\documentclass{article}
 
\usepackage[a4paper,bindingoffset=0.2in,
             left=0.9in,right=0.9in,top=1in,bottom=1in,
             footskip=.25in]{geometry}

\usepackage[utf8]{inputenc}
\usepackage{pgfplots}
\DeclareUnicodeCharacter{2212}{−}
\usepgfplotslibrary{groupplots,dateplot}
\usetikzlibrary{patterns,shapes.arrows}
\pgfplotsset{compat=newest}
\usetikzlibrary{arrows,matrix,positioning}
\usepackage[normalem]{ulem}
\usepgfplotslibrary{fillbetween}
\usepackage{hyperref}
\usepackage{algorithm}
\usepackage{algpseudocode}
\usepackage{pifont}
\usepackage{siunitx}  
\usepackage[authoryear]{natbib}
\usepackage[normalem]{ulem}
\usepackage{amssymb}
\usepackage{amsfonts} 
\usepackage{amsmath} 
\usepackage{url}
\usepackage{cite}
\usepackage{multirow, booktabs}
\usepackage{color}
\usepackage{tcolorbox}
\usepackage{todonotes}
\usepackage{setspace}
\usepackage{booktabs} 
\usepackage{dcolumn}
\usepackage{siunitx}  
\newcolumntype{Y}{D..{4.4}}

\newlength{\bibitemsep}
\newlength{\bibparskip}
\let\oldthebibliography\thebibliography
\renewcommand\thebibliography[1]{%
  \oldthebibliography{#1}%
  \setlength{\parskip}{\bibitemsep}%
  \setlength{\itemsep}{\bibparskip}%
}

\newcommand{\bs}{\boldsymbol}
\newcommand{\mbf}{\mathbf}

\newcommand{\mcl}{\mathcal}
 
\usepackage{xstring}
\def\alphabet{abcdefghijklmnopqrstuvwxyzABCDEFGHIJKLMNOPQRST123456789}
\renewcommand{\vec}[1]{
\IfSubStr{\alphabet}{#1}{
\ensuremath{\mathbf{\MakeLowercase{#1}}}
}{
\ensuremath{\boldsymbol{\MakeLowercase{#1}}}
}
}
\newcommand{\mat}[1]{
\IfSubStr{\alphabet}{#1}{
\ensuremath{\mathbf{\MakeUppercase{#1}}}
}{
\ensuremath{\boldsymbol{\MakeUppercase{#1}}}
}
}

\def\R{\mathbb R}
\def\C{\mathbb C}

\newcommand*{\card}[1]{\left|#1\right|}

\newcommand{\m}[1]{{\mathrm{#1}}}
\newcommand{\um}[1]{_{\mathrm{#1}}}

\newcommand{\nrn}{n_{\mathrm{r}}}

\newcommand{\affiliation}[1]{\def\@affil{#1}} 

\usepackage{acronym}
\acrodef{HO}[HO]{harmonic oscillator}
\acrodef{SHO}[SHO]{spherical ha1rmonic oscillator}
\acrodef{CHO}[CHO]{circular harmonic oscillator}
\acrodef{PDE}[PDE]{partial differential equation}
\acrodef{ODE}[ODE]{ordinary differential equation}
\acrodef{FT}[FT]{Fourier transform}
\acrodef{WDD}{Wigner Distribution Deconvolution}
\acrodef{STEM}{Scanning Transmission Electron Microscopy}
\acrodef{SSB}{Single Side Band}
\acrodef{COM}{Center of Mass}
\acrodef{WSe$_2$}{Tungsten diselenide}
\acrodef{SrTiO$_3$}{Strontiumtitanat}
\acrodef{UDF}{User Defined Function}
\acrodef{ROI}{Region of Interest}

\begin{document}  
\title{Wigner Distribution Deconvolution Adaptation for Live Ptychography Reconstruction}
\author{\normalsize Arya Bangun\thanks{Ernst Ruska-Centre for Microscopy and Spectroscopy with Electrons, Forschungszentrum J\"ulich, 52425 J\"ulich, Germany},
  Paul F.~Baumeister\thanks{J\"ulich Supercomputing Centre, Forschungszentrum J\"ulich, 52425 J\"ulich, Germany},
  Alexander Clausen\footnotemark[1] ,
  Dieter Weber\footnotemark[1],
  Rafal E. Dunin-Borkowski\footnotemark[1]
  }

\maketitle 
\begin{abstract} 
We propose a modification of  \ac{WDD} to support live processing ptychography. Live processing allows to reconstruct and display the specimen transfer function gradually while diffraction patterns are acquired. For this purpose we reformulate \ac{WDD} and apply a dimensionality reduction technique that reduces memory consumption and increases processing speed. We show numerically that this approach maintains the reconstruction quality of specimen transfer functions as well as reduces computational complexity during acquisition processes. \textcolor{black}{Although we only present the reconstruction for \ac{STEM}  datasets, in general,  the live processing algorithm we present in this paper can be applied to real-time ptychographic reconstruction for different fields of application.}
\end{abstract}

\section{Introduction} \label{sec:intro}
\begin{figure}[htb!]
\centering
\includegraphics[scale=0.67]{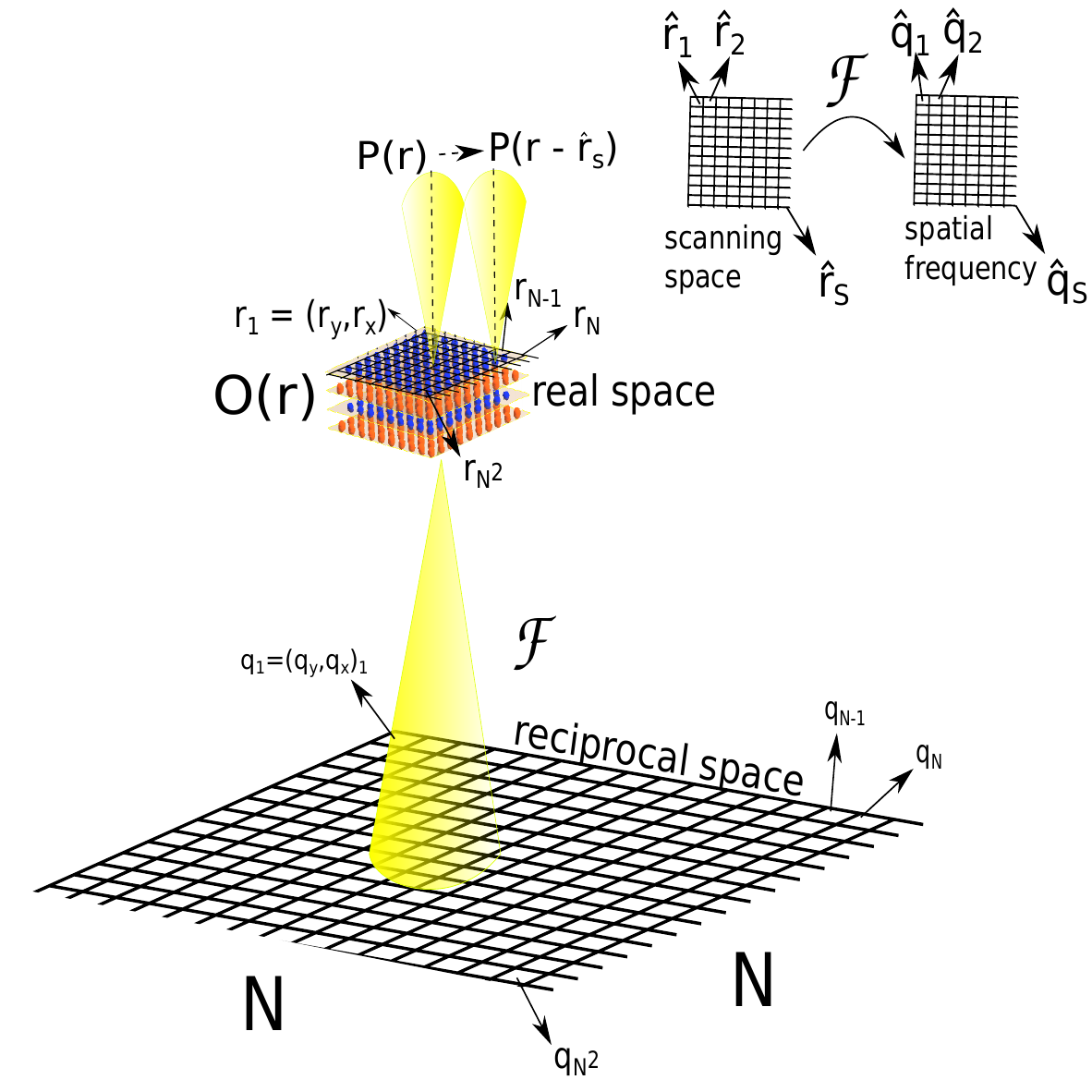}
\caption{Four-dimensional \ac{STEM} acquisition for a scanning point $s \in [S]$ with definition of real space grid $\mathbf{r} = (r_y, r_x) \in \mathbb{R}^2$, reciprocal space grid $\mathbf{q} = (q_y, q_x) \in \mathbb{R}^2$, scanning grid $\hat{\mbf r} = (\hat{r}_y, \hat{r}_x) \in \mathbb{R}^2$ and spatial frequency grid $\hat{\mbf q} = (\hat{q}_y, \hat{q}_x) \in \mathbb{R}^2$.}
\label{Fig:Ptycho}
\end{figure}

%
Four-dimensional \ac{STEM} is an experimental modality where a wide range of computational methods can extract information on the specimen
and reduce the acquired data for human interaction \citep{Ophus2019}. The acquisition schema of four-dimensional \ac{STEM}
is shown in Figure \ref{Fig:Ptycho}. Acquiring such comprehensive data and applying computational ana\-lysis and reconstruction workflows allows observation of material properties by electron microscopy that are not accessible with simple detectors and signal processing me\-thods \citep{Ophus2019}. Sin\-ce it generates large amounts of data \citep{spurgeon2021towards}, making algorithms and implementations efficient in their use of computer memory and processing time requires special attention.

If a computational method to process the recorded data is only implemented for offline use, microscopists have
to acquire data relying only on simple contrast methods or even without any feedback at all. An implementation for live processing, in contrast, allows interactive use of the microscope based on advanced computational contrast mechanisms, monitoring the acquisition process, evaluating data quality, or automatically controlling the instrument in a closed loop. This requires suitable interfaces to the microscope to receive a live data stream as well as implementations that are capable to process data gradually while they arrive \citep{nord2020fast}.

Ptychography \citep{Hoppe1969,
Hoppe1969a, Hoppe1969b} can be used to extract a quantitative object transfer function for a specimen using 4D \ac{STEM} data. It takes advantage of  information from local overlap of the illuminated regions. Recent years have seen a widespread increase in the development of ptychography algorithms by different approaches, such as inspired by the classical alternating projection method, i.e., PIE reconstruction algorithm  \citep{rodenburg2004phase,maiden2009improved}, other optimization-based approaches  \citep{bostan2018accelerated, wen2012alternating}, as well as direct
methods. \ac{SSB} \citep{Rodenburg1993SSB, pennycook2015efficient,yang2015efficient} and \ac{WDD} \citep{rodenburg1992theory,li2014ptychographic,yang2016simultaneous} are examples of direct ptychography methods that extract the relevant information in a sequential processing flow, as opposed to iterative methods that optimize the object transfer function in a loop over the input data.

With \ac{SSB} as an
example, Strauch et al. \citep{Strauch2021} showed previously that direct, linear methods are particularly suitable for live processing since the result can always be expressed as the sum of partial results from processing subsets
of the input data. However, reconstruction with \ac{SSB} relies on the weak phase object approximation. Compared to \ac{SSB}, \ac{WDD} does not have this limitation \citep{yang2017electron}. At the same time, the data reduction in \ac{WDD} is a linear function of the input data like in \ac{SSB}, meaning it is a good candidate for live ptychography. Strong binning is used as dimensionality reduction method in \citep{pelz2021real} to reduce the processing time for \ac{SSB}
after an acquisition is completed.

\subsection{Related Works}

A real-time phase reconstruction approach based on integrated \ac{COM} is proposed in \citep{yu2021real}. That method does not require storing the entire four-dimensional dataset in memory and reconstructs the phase from strongly reduced \ac{COM} information instead of full diffraction patterns. The \ac{COM} can be computed efficiently from diffraction data so that this algorithm can reconstruct large-scale data. The authors coined the real-time approach as ri\ac{COM}.

\subsection{Summary of Contributions}
Here we demonstrate live reconstruction using the eigenfunctions of a harmonic oscillator instead of binning as a base for dimensionality reduction. In addition to data reduction, this transformation can replace the Fourier transform in several steps of the \ac{WDD} method while retaining a reconstruction that is very similar to a result without dimensionality reduction. Furthermore, by building on the matrix representation of the discrete Fourier transform, we can process the intensity data and reconstruct the specimen transfer function gradually from subsets of the input data in a streaming fashion. As a benchmark, we compare the result of Live \ac{WDD} with the implementation of \ac{WDD} in PyPtychoSTEM \citep{PyPtychoSTEM} as well as Live \ac{SSB} \citep{Strauch2021}.

The codes used and to reproduce the result in this paper are available at the URL below:
\begin{center}
  \textcolor{black}{\url{https://github.com/Ptychography-4-0/LiveWDD}} 
\end{center}

\subsection{Notations}
We provide notations used throughout this article. Vectors are written in bold
small-cap letters $\mathbf{x} \in \C^L$ and matrices are written as a bold
big-cap letter $\mathbf{A} \in \C^{K \times L}$ for a complex field $\C$ and for
a real field $\R$.  A matrix can also be written by indexing its elements
\[\mathbf{A} = \left(a_{k\ell}\right), \quad \text{where} \quad k \in [K], \ell
\in [L]. \] The set of integers is written as $[N] := \{1,2,\hdots,N\}$ and
calligraphic letters are used to define functions $\mathcal{A} : \C \rightarrow
\C$. Specifically, we denote the discrete two-dimensional Fourier transform by
$\mathcal{F}$. For both matrices and vectors, the notation $\circ$ is used to
represent an element-wise product, also called Hadamard product. $\mbf{A}^T$ is used to
denote the transpose of a matrix $\mbf{A}$. The notation $\text{vec}: \C^{N \times N}
\rightarrow \C^{N^2}$ is an operator that vectorizes a matrix, and
$\text{mat}:\C^{N^2} \rightarrow \C^{N \times N}$ constructs a matrix from a vector.
The complex conjugate is indicated by a bar. For a matrix $\overline{\mbf X}$, it
is the element-wise conjugate.  The convolution and Kronecker operator are denoted by
$\circledast$ and $\otimes$, respectively. The partial derivative is given by
$\partial$, and $\nabla$ is the nabla or vector differential operator.

\section{Wigner Distribution Deconvolution}\label{Sec2:WDD}

In this section, we introduce the matrix representation of the \ac{WDD} method developed from its original formulation as in \citep{rodenburg1992theory} and an Open Source implementation from \citep{PyPtychoSTEM}. In order to visualize the definition for different spaces used in this article we refer to Figure \ref{Fig:Ptycho}.
The specimen transfer function is denoted by a matrix $\mbf{O} \in \C^{N \times N}$ where values at the row and column indices $(i, j)$ define the object function at a position $\mbf{r} \in [N^2]$ in the specimen plane. Similarly, values of the illuminating probe $p(\mbf r)$ in the specimen plane at position $\mbf{r}$ without shifting can be written in matrix form as $\mbf{P} \in \C^{N \times N}$.

For all shifting coordinates in the set of flattened scan position coordinates, $\hat{\mbf{r}_s} \in \{\hat{\mbf{r}}_1,\hat{\mbf{r}}_2,\dots,\hat{\mbf{r}}_S \}$, we write the shifted matrix probe as $\mbf{P}_s$ for $s \in [S]$, where the $S = S_y \times S_x$ is the set of scanning points. Consequently, the matrix for a shifted matrix probe is given by
$$
\left(p_{ij} \right)_s = p(\mbf r - \hat{\mbf r}_s), \quad \text{where} \quad  i,j \in [N],
$$
Here we define the object, probe, diffraction patterns and the scanning points on equi-spaced rectangular grids. Hen\-ce, the usual implementation of the discrete Fourier transform can be used. The intensity of the diffraction pattern at each scanning point can be written as
\begin{equation}
    \label{eq:forward}
\mbf{I}_s = \card{\mcl{F}_{\mbf{r}}\left(\mbf{P}_s \circ \mbf{O} \right)}^2 \quad \text{for} \quad s \in [S].
\end{equation}
Similar to the real space coordinate, each pattern  $\mbf{I}_s$  for scan position $s$ is indexed by flattened reciprocal space coordinates $\mbf{q}_s \in \{\mbf{q}_1,\mbf{q}_2,\dots,\mbf{q}_{N^2} \}$, where $N^2$ denotes the pair index in the two-dimensio\-nal grid in reciprocal space. It should be noted that the complete set of diffraction patterns can be written as flattened scanning points index $\mbf{I} \in \R^{S \times N \times N}$ or as a four-dimensional tensor $\mbf{I} \in \R^{S_y \times S_x \times N \times N}$ indexed by the two-dimensional scan position $\hat{\mbf{r}}$ and the two-dimensional reciprocal space coordinate $\mbf{q}$, corresponding to the diffraction angle. 

The two-dimensional Fourier transform with respect to the object and probe grid coordinates $\mbf{r}$ is written as $\mcl{F}_{\mbf{r}}$, with respect to the scan coordinate $\hat{\mbf{r}}$ as $\mcl{F}_{\hat{\mbf{r}}}$, and the inverse transforms consequently as $\mcl{F}_{\mbf{q}}^{-1}$ and $\mcl{F}_{\hat{{\mbf{q}}}}^{-1}$. The \ac{WDD} algorithm estimates the object from the intensity of diffraction patterns and an estimate of the probe. Picking up from \citep[eq.8]{rodenburg1992theory} where a relation between the object's and probe's Wigner distributions $\mbf{W}_O \in \C^{S_y \times S_x \times N \times N}$ resp. $\mbf{W}_P \in \C^{S_y \times S_x \times N \times N}$ is shown:
\begin{equation}
    \label{eq:deconv_mat}
    \mbf{W}_O = \frac{\mcl{F}_{\mbf{{q}}}^{-1}\left( \mcl{F}_{\hat{{\mbf r}}}\left( \mbf{I} \right)\right) \overline{\mbf{W}_{P}}}{\card{\mbf{W}_{P}}^2 + \epsilon} 
\end{equation}
with $ \mbf{W}^v_{P} = \mcl{F}_\mbf{q}^{-1}\left(\mcl{F}_\mbf{r}\left(\mbf{P}\right) \overline{\mcl{F}_\mbf{r}\left(\mbf{P}\right)}\big\vert_{\left(\mbf{q} + \hat{\mbf{q}}_v\right)}\right)$, where $\mcl{F}_{\mbf{r}}\left(\mbf{P}\right)\big\vert_{\left(\mbf q + \hat{\mbf{q}}_v\right)}$ denotes a shift in reciprocal space of the Fourier-transform\-ed probe for specific index $v$.

By inserting an estimate of the probe $\mbf{\widetilde P}$ and calculating the probe's Wigner distribution $\mbf{\widetilde{W}}_{P}$, one can estimate the object's Wigner distribution $\mbf{\widetilde{W}}_{O}$. Following \citep{rodenburg1992theory}, an estimate for the object $\mbf{\widetilde O}$ can be obtained from $\mbf{\widetilde{W}}_{O}$ as
\begin{equation}
    \label{eq:extract}
    \mbf{\widetilde{O}} = \frac{
        \mcl{F}_{\hat{\mbf{q}}}^{-1}\left(\mcl{F}_{\mbf r}\left(\mbf{\widetilde{W}}_{O}\right)\bigg\vert_{\left(\mbf{q} = 0\right)}\right)
    }{
        \sqrt{\mcl{F}_\mbf{r}\left(\mbf{\widetilde{W}}_{O}\right)\bigg\vert_{\left(\hat{\mbf{q}} = 0, \mbf{q} = 0\right)}}
    } \in \C^{S_y \times S_x},
\end{equation} 
with $\mbf{q} = 0$ and $\left(\hat{\mbf{q}} = 0, \mbf{q} = 0\right)$ denoting extraction of a subset at the specified coordinates.
\begin{figure*}[ht!]
    \centering
    \includegraphics[scale=0.55]{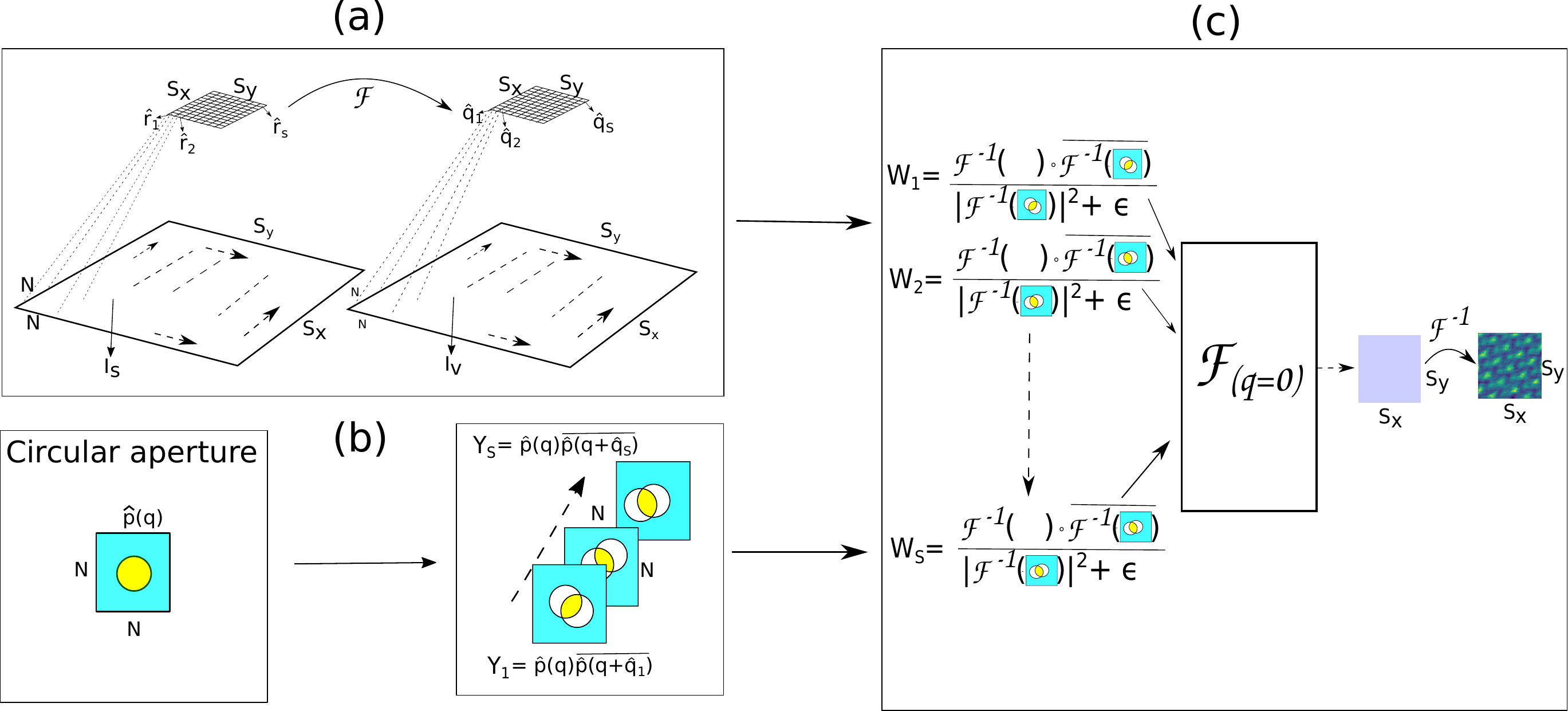}
    \caption{\textcolor{black}{The schematic diagram for classical \ac{WDD}. (a). The calculation of spatial frequencies $\hat{\mbf{q}}_v$ of a four-dimensional \ac{STEM} diffraction pattern dataset is given by applying two-dimensional Fourier transform in the real space scan coordinate $\hat{\mbf{r}}_s$, (b). For each spatial frequency the autocorrelation of the circular aperture is calculated, i.e., initial probe in reciprocal space, (c). Apply a two-dimensional inverse Fourier transform for each diffraction pattern dataset on each spatial frequency and the probe autocorrelation and use it for the deconvolution process by using a Wiener filter. Afterward, apply Fourier transform and extract for reciprocal space $\mbf{q} = 0$ to get an image on spatial frequency before estimating the object in real space scan coordinate with a two-dimensional inverse Fourier transform.
    }}
    \label{Fig:schematic_WDD}
\end{figure*}
A visualization of the \ac{WDD} algorithm is depicted in Figure \ref{Fig:schematic_WDD}.

Calculating the deconvolution in \eqref{eq:deconv_mat} may consume lar\-ge amounts of memory for typical four-dimensional \ac{STEM} data if implemented naively following the equations above since a Fourier transform of the original data, as well as $\mbf{W}_{P}$, which has the same size as the input data, might be instantiated at the same time. Furthermore, they are complex-valued and may require higher numerical precision than the input data. Additionally, this algorithm works on entire datasets, preventing a direct implementation of live processing.

To address the computational complexity and allow live processing we introduce a dimensionality reduction to reduce the size of the input data and $\mbf{W}_{P}$, and rearrange the \ac{WDD} algorithm. The modification allows us to process the intensity data sequentially and update the reconstruction as the scan proceeds to acquire new intensity data.

\section{Validation}

Since ptychography is a quantitative reconstruction technique, any implementation should demonstrate that it is correct, i.e., that it reconstructs arbitrary object functions and/or illuminations quantitatively within its inherent limitations. Simulated datasets of a crystalline specimen resemble the real-world data that ptychography is used on in electron microscopy, but they often have very high symmetry and have no natural orientation. That means errors such as a rolled, phase-reversed or inverted reconstruction, or multiplication with a factor may not be obvious. Such errors can, for example, be caused by mixing Fourier transform and inverse Fourier transform, by adding or omitting a complex conjugation, or by incorrect use of FFT shift resp. inverse FFT shift. For this reason the implementations used in this paper were validated with a simple procedurally generated asymmetric test image. It is clearly recognizable starting at $25\times25$~px and contains a wrapped phase ramp at an odd angle that creates a characteristic single spot in the Fourier transform, allowing to visually confirm the correct relation between real and reciprocal space.

\begin{figure}[!htb]
    \centering
    \includegraphics[scale=0.45]{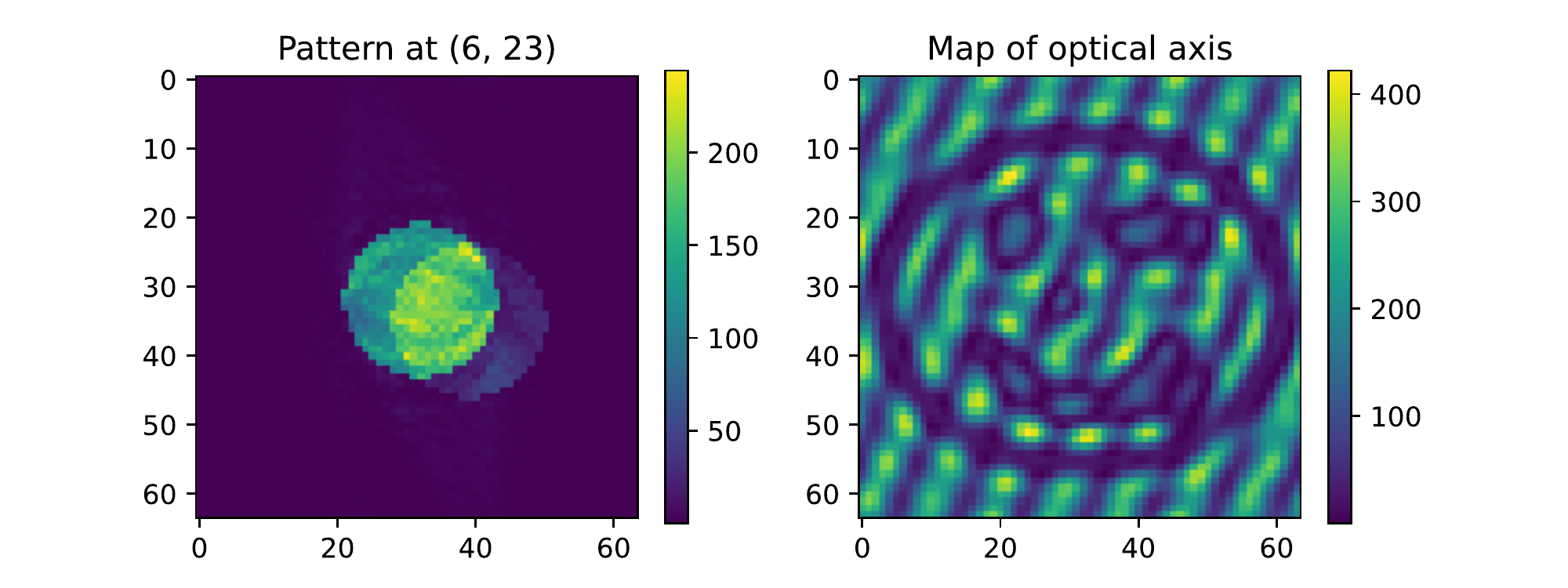}
    \caption{Sample diffraction pattern and projection along the optical axis from the synthetic test dataset that is used for validation.}
    \label{fig:sample} 
\end{figure}

A test dataset was created from this test object using multiplication with a synthetic illumination rolled to the scan position, inverse FFT shift, Fourier transform and FFT shift. The illumination was calculated from a synthetic circular aperture with value 17 to catch scaling issues that may not be apparent if the value 1 was used. The illumination function in amplitude and phase as well as the calculated amplitude, phase and intensity from the forward simulation were visually inspected to conform with the expected values. In particular, the simulated diffraction patterns contain a shifted replica of the illuminating aperture at the expected position as a signature of the wrapped phase ramp. A virtual bright field image of the simulated dataset confirmed the correct spatial arrangement of the diffraction patterns. See Figure \ref{fig:sample} for a sample diffraction pattern and a virtual bright field image.

\begin{figure}[!htb]
    \centering
    \includegraphics[scale=0.35]{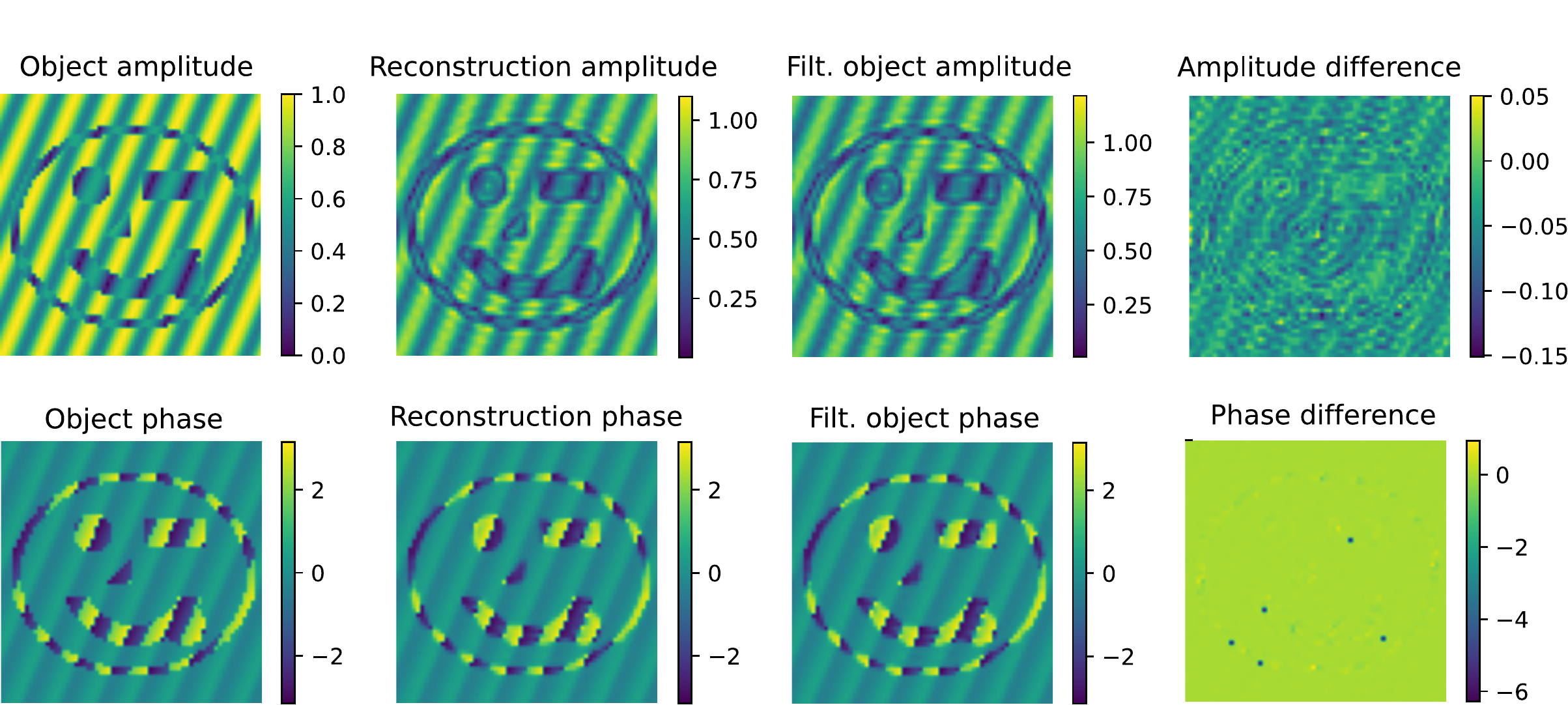}
    \caption{Validation with a procedurally-generated synthetic test dataset: Comparing original object, reconstruction, approximate of the expected result calculated by bandpass-filtering the object with twice the aperture size, and difference between reconstruction and approximate expected value.}
    \label{fig:validation} 
\end{figure}

The \ac{WDD} implementation was then confirmed to reconstruct the object quantitatively in amplitude and phase, taking the band pass filtering of \ac{WDD} with twice the semiconvergence angle of the aperture into account (Figure \ref{fig:validation}). The correct scale for the illumination of real-world microscopy data can be derived from a vacuum reference, i.e. a scan with the same parameters but without the specimen. This enables quantitative reconstruction of both amplitude and phase of the object function. A Jupyter notebook with the validation is available at \url{https://github.com/Ptychography-4-0/LiveWDD}.

\section{Circular Harmonic Oscillator}\label{sec:sho}
%
 In this section we present details of the dimensionality reduction method used.

The rate-limiting step in conventional \ac{WDD} is the deconvolution of a massive dataset. Therefore, one can improve the performance by projecting the dataset onto a space with lower dimension while retaining the essential information, and performing the resource-intensive steps in this reduced representation. Cropping and binning are simple examples of such projections from higher to lower dimension.

A basis of eigenfunctions of the harmonic oscillator has beneficial properties in this application that are detailed in the following sections. We start first by defining the harmonic oscillator. We provide a brief introduction to this topic and refer the interested readers to the literature \citep{Sakurai:1167961}.
\subsection{One-Dimensional Harmonic Oscillator}
\begin{figure}[!htb]

  \centering
	\includegraphics[scale=0.3]{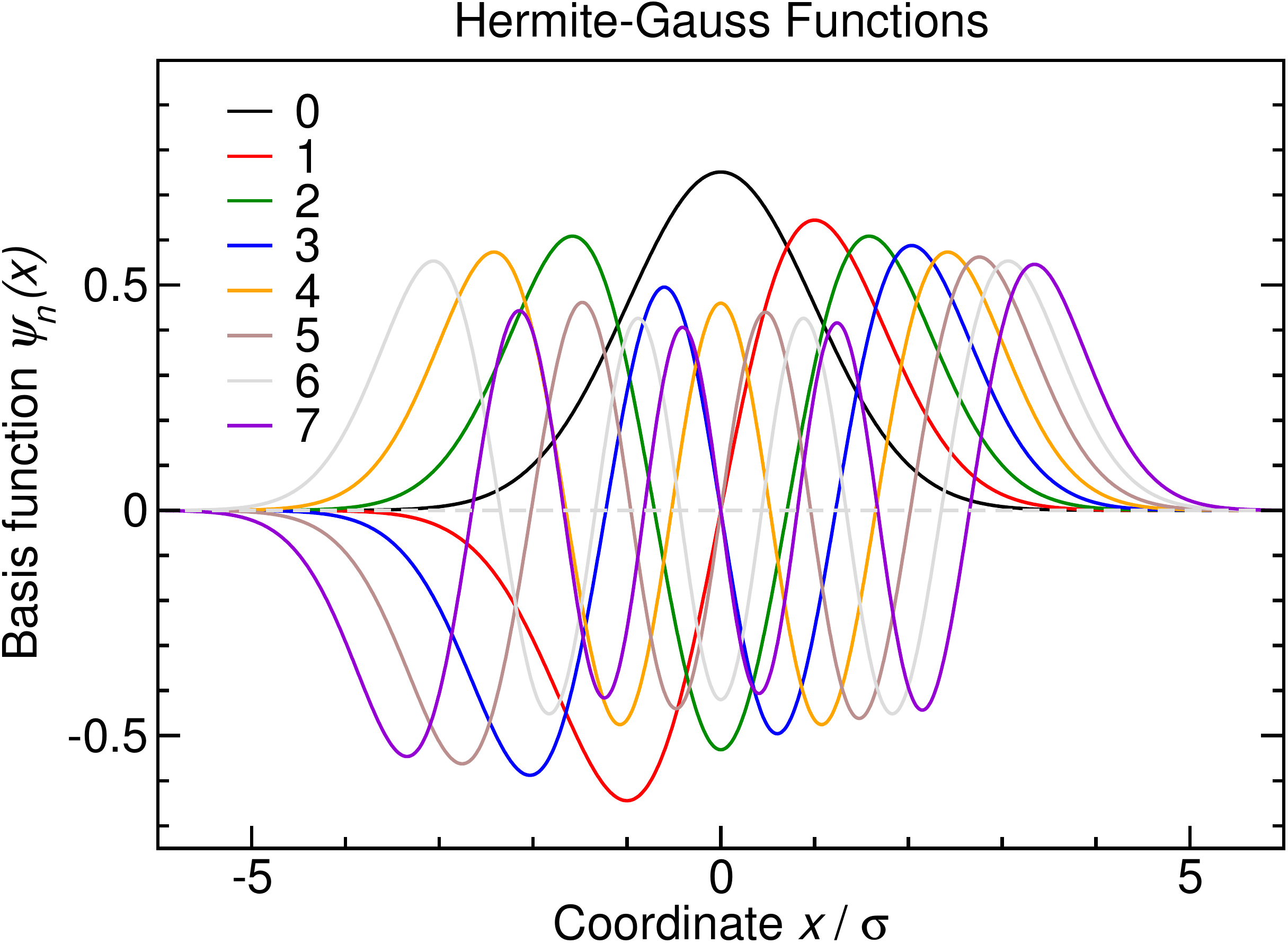} 
 
  \caption{
  Normalized Hermite-Gauss functions up to $n\um{max} = 7$. 
  All functions are either even ($\psi(-x) = \psi(x)$) or odd ($\psi(-x) = -\psi(x)$), depending on the parity of $n$.
  }
   \label{fig:hermite_gauss_function}
\end{figure}

The quantum-mechanical \ac{HO} problem is described by the Hamiltonian operator
\begin{equation}
  \hat H^{[1D]}_{\sigma} = -\frac{\partial_x^2}{2} + \frac{x^2}{2 \sigma^4} \text{.}
  \label{eqn:HO-Hamiltonian}
\end{equation}
Hartree atomic units are used in this section.
Here, $\sigma$ is a length scale parameter that also fixes the scale of the eigenvalues $E^{[1D]}_{n}(\sigma)$ 
of the operator $\hat H^{[1D]}_{\sigma}$, the so called eigenenergies.
The eigenvalue problem reads
\begin{equation}
  \hat H^{[1D]}\psi^{[1D]}_{n}(x) = E^{[1D]}_{n} \psi^{[1D]}_{n}(x)\text{,} \   n \in \mathbb N_0 \text{.}
  \label{eqn:HO-eigenvalueproblem}
\end{equation}
Then, the energy eigenvalues are
\begin{equation}
  E^{[1D]}_{n}(\sigma) = \sigma^{-2} \left( n + \frac 12 \right)
  \label{eqn:HO-eigenenergy}
\end{equation}
and the corresponding \ac{HO} eigenfunctions are
\begin{equation}
  \psi^{[1D]}_{n}(x) = H_n\left(\frac{x}{\sigma}\right) \   \exp\left( -\frac{x^2}{2 \sigma^2} \right) 
  \label{eqn:HO-eigenfunction}
\end{equation}
with the Hermite polynomials $H_n$.
It can be seen that these eigenfunctions are Hermite polynomials weighted by a Gaussian envelope function.
For the usage as a basis, a proper $L_2$-normalization is necessary. \textcolor{black}{The normalized Hermite polynomials for several degrees are presented in Figure \ref{fig:hermite_gauss_function}.}

\subsection{Two-Dimensional Isotropic Harmonic Oscillator}

The eigenfunctions of the quantum-mechanical 
two-dime\-nsional isotropic harmonic oscillator, in the following we will refer to it as \ac{CHO}, can be written as Cartesian product of two
\ac{HO} eigenfunctions. The Hamiltonian
\begin{equation}
  \hat H^{[2D]}_{\sigma} = -\frac{\vec \nabla^2}{2} + \frac{\vec r^2}{2 \sigma^{4}} 
  \label{eqn:CHO-Hamiltonian}
\end{equation}
has the eigenfunctions
\begin{equation}
  \psi^{[2D]}_{n_x n_y}(x,y) = \psi^{[1D]}_{n_x}(x/\sigma)
                           \   \psi^{[1D]}_{n_y}(y/\sigma) 
  \label{eqn:CHO-eigenfunction}
\end{equation}
and the eigenenergies
\begin{equation}
  E^{[2D]}_{n_x n_y}(\sigma) = E^{[1D]}_{n_x}(\sigma) + E^{[1D]}_{n_y}(\sigma) = \sigma^{-2} \left( n_x + n_y + 1 \right) 
  \label{eqn:CHO-eigenenergy}
\end{equation}

\subsection{Special property of the basis}

The \ac{HO} eigenfunctions with $\sigma = 1$ are also eigenfunctions of the \ac{FT} operator, i.e.~the Fourier transform of $\psi^{[1D]}_{n}(x)$ is again a Gauss-Hermite function: 
\begin{equation}
    \mathcal{F}\left(\psi^{[1D]}_{n}(x) \right) = \imath^n \psi^{[1D]}_{n}(q)
    \label{eqn:eigenfunction-of-FT},
\end{equation}
for $q$ being the reciprocal space coordinate adjoint to $x$.
For a non-unity spread, i.e.~any $\sigma > 0$, the \ac{FT} produces a spread of $1/\sigma$ in reciprocal space.

Furthermore, transforming a function by this eigenfunction changes a convolution into multiplication similar to a Fourier transform, which is discussed in \citep[Theorem 4.1]{glaeske1983convolution}. That means transforming into this basis can replace Fourier transforms in the \ac{WDD} algorithm.

\section{Dimensionality Reduction for WDD} 
\label{sec:dr_wdd}
The Hermite-Gauss functions defined in \eqref{eqn:HO-eigenfunction} can be used as a basis for dimensionality reduction of \ac{STEM} datasets. We first introduce the matrix notation from sampled Her\-mite-Gauss function before presenting the procedure to reduce the dimension of the data. Secondly, the transformation will be presented as well as a numerical example for the transformation.

The sampling grid $x$ and scaling factor $\sigma$ in \eqref{eqn:HO-eigenfunction} should be adapted so that the basis is centered with respect to the diffraction patterns and scaled to cover the area relevant for WDD, i.e. the area illuminated by the primary beam.
 
We can construct a matrix from sampled Hermite-Gauss functions as presented below
$$
\small
\bs{\psi}_x = 
\begin{pmatrix}
  \psi^{[1D]}_{n_1}(x_1) \quad \psi^{[1D]}_{n_2}(x_1)&\dots& \psi^{[1D]}_{n_L}(x_1) \\
    \psi^{[1D]}_{n_1}(x_2) \quad \psi^{[1D]}_{n_2}(x_2)&\dots& \psi^{[1D]}_{n_L}(x_2) \\
  \vdots&  \dots&\vdots\\
    \psi^{[1D]}_{n_1}(x_N) \quad \psi^{[1D]}_{n_2}(x_N)&\dots& \psi^{[1D]}_{n_L}(x_N) \\
\end{pmatrix} \in \R^{N \times L},
$$
where we construct a sampling grid with respect to the center of mass in the $x$ direction. After introducing the matrix representation of the Hermite-Gaussian function, we can define the dimensionality reduction by using this matrix. For a $\mbf{A} \in \R^{N \times N}$ the projection to the lower dimension using Hermite-Gauss functions can be expressed as
\begin{equation}
\mcl{H}\left(\mbf{A}\right) := \bs{\psi}_x^T\mbf{A}^T \bs{\psi}_y \in \R^{L \times L}
\end{equation}
This function $\mcl{H}: \C^{N \times N} \rightarrow \C^{L \times L}$ where $L \ll N$ maps from the data with dimension $N \times N$ to a lower dimension $L \times L$. 


In contrast to the original two-dimensional discrete Fourier transformation that preserves the dimension of the dataset, this transformation allows flexibility to reduce the dimension. Hence, we can apply this dimensional reduction technique in the deconvolution step in \ac{WDD} as presented below
\begin{equation}
\begin{aligned}
\label{eq:hermite_conv}
\mcl{H}\left(\mbf{I}_v \right) &= \mcl{H}\left(\mbf{X}_v \circledast \mbf{Y}_v\right) \quad \text{for} \quad v \in [S]\\
&= \mcl{H}\left(\mbf{X}_v\right) \mcl{H}\left(\mbf{Y}_v\right),
\end{aligned}
\end{equation}
where now the dimension is reduced from $N \times N$ to $L \times L$ with $L \ll N$. The variable $\mathbf{I}_v$ is the intensity of diffraction pattern of specific spatial frequency coordinate $v \in [S]$. Variables $\mbf{Y}_v$ and $\mbf{X}_v$ represent the autocorrelation of the illuminating probe and the autocorrelation of the specimen transfer function, respectively. In parallel with classical \ac{WDD}, the $\mcl{H}\left(\mbf{Y}_v\right)$ and $\mcl{H}\left(\mbf{X}_v\right)$ represent functions at specific spatial frequency $v$ that have similar properties to the Wigner distribution function of the probe and object in conventional \ac{WDD}.

\begin{figure}[!htb]
\centering
     \scalebox{0.3}{\pgfplotsset{every tick label/.append style={font=\Huge}}
\begin{tikzpicture}
\begin{groupplot}[group style={group size=3 by 1,horizontal sep=8.5em},width = 0.5*\textwidth, height=(0.5)*\textwidth]
\nextgroupplot[
colorbar horizontal,
colormap/viridis,
colorbar style={xtick={0,810487},scaled x ticks=false,xticklabels={0, $8.1e5$}},
hide x axis,
hide y axis,
point meta max=810487,
point meta min=0,
tick align=outside,
tick pos=left,
x grid style={white!69.0196078431373!black},
xmin=-0.5, xmax=255.5,
xtick style={color=black},
y dir=reverse,
y grid style={white!69.0196078431373!black},
ymin=-0.5, ymax=255.5,
ytick style={color=black}
]
\addplot graphics [includegraphics cmd=\pgfimage,xmin=-0.5, xmax=255.5, ymin=255.5, ymax=-0.5] {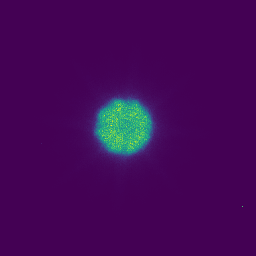};

\nextgroupplot[
colorbar horizontal,
colorbar style={xtick={-556567.375,14353222},scaled x ticks=false,xticklabels={$-5.5e5$, $1.43e7$}},
colormap/viridis,
hide x axis,
hide y axis,
point meta max=14353222,
point meta min=-556567.375,
tick align=outside,
tick pos=left,
x grid style={white!69.0196078431373!black},
xmin=-0.5, xmax=15.5,
xtick style={color=black},
y dir=reverse,
y grid style={white!69.0196078431373!black},
ymin=-0.5, ymax=15.5,
ytick style={color=black}
]
\addplot graphics [includegraphics cmd=\pgfimage,xmin=-0.5, xmax=15.5, ymin=15.5, ymax=-0.5] {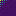};

\nextgroupplot[
colorbar horizontal,
colorbar style={xtick={0,810487},scaled x ticks=false,xticklabels={0, $8.1e5$}},
colormap/viridis,
hide x axis,
hide y axis,
point meta max=810487,
point meta min=0,
tick align=outside,
tick pos=left,
x grid style={white!69.0196078431373!black},
xmin=-0.5, xmax=255.5,
xtick style={color=black},
y dir=reverse,
y grid style={white!69.0196078431373!black},
ymin=-0.5, ymax=255.5,
ytick style={color=black}
]
\addplot graphics [includegraphics cmd=\pgfimage,xmin=-0.5, xmax=255.5, ymin=255.5, ymax=-0.5] {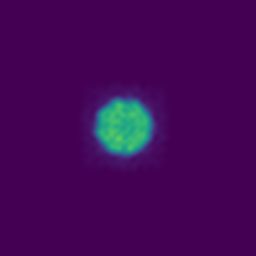};
\end{groupplot}
 
\end{tikzpicture}}  

\caption{Dimensionality reduction of PACBED data from \ac{SrTiO$_3$} dataset  \citep{strauch2021high}. (a) PACBED data, (b) Lower Dimension $(16 \times 16)$, (c) Reconstructed PACBED}
\label{Fig:Comp_IDP}
\end{figure}

Transformation to lower dimension can be seen in Figure \ref{Fig:Comp_IDP}, where we have reduced from dimension $256 \times 256$ to $16 \times 16$. Since we apply dimensionality reduction to our four-dimensional \ac{STEM} data, we also have to apply it to the autocorrelation of the illuminating probe to proceed with the deconvolution process, as shown in \eqref{eq:hermite_conv}. The transformation from convolution into multiplication by applying the operator $\mcl{H}$ is derived from the properties of Hermite-Gaussian functions applied to the convolution as introduced in the previous section. Compared to conventional \ac{WDD}, the deconvolution can be performed in a lower dimension and in a different space with similar properties, and the rest follows similar to \ac{WDD}.

\section{Live Processing \ac{WDD}}\label{Sec: Live_Proc}
A conventional \ac{WDD} implementation following the procedure in Section \ref{Sec2:WDD} wo\-rks on an entire dataset, which is not suitable for true live processing. A WDD implementation for live processing should update the estimate for the object transfer function gradually by processing individual diffraction patterns as they arrive from the acquisition process. In \ac{WDD} a Fourier transform is applied over the scan position coordinates, which is usually performed at once with an FFT in a conventional implementation, hence requiring the entire dataset.

Furthermore, live processing should be fast enough to keep up with typical detector speeds that are in the kHz range for 4D \ac{STEM}. The strategy to support live processing with \ac{WDD} consists of three steps:
\begin{enumerate}
  \item Implementation of the Fourier transform over the scan position coordinates using multiplication with a partial DFT matrix instead of FFT.
  \item Separation and pre-computation of variables that can be computed independent of the diffraction patterns, in this case the Wigner distribution function of the probe and Wiener filter in \eqref{eq:deconv_mat}.
  \item Dimensionality reduction to reduce number of individual computations.
\end{enumerate}

\subsection{Fourier transform}
We begin with a quick introduction of implementing a Fourier transform with a DFT matrix. 

A one-dimensional Fourier transform can be implemented by constructing a Fourier matrix taken from sampled Fourier basis as follows \citep[eq.5.44]{jain1989fundamentals},

$$
\mathbf{F} = \frac{1}{\sqrt{N}}\begin{pmatrix}
  e^{\frac{-i2\pi f_1x_1}{N}} \quad e^{\frac{-i2\pi f_1x_2}{N}}&\dots& e^{\frac{-i2\pi f_1x_N}{N}} \\
  e^{\frac{-i2\pi f_2x_1}{N}} \quad e^{\frac{-i2\pi f_2x_2}{N}}&\dots& e^{\frac{-i2\pi f_2x_N}{N}} \\
  \vdots&  \dots&\vdots\\
  e^{\frac{-i2\pi f_Nx_1}{N}} \quad e^{\frac{-i2\pi f_Nx_2}{N}}&\dots& e^{\frac{-i2\pi f_Nx_N} {N}},\\
\end{pmatrix} \in \C^{N \times N},
$$
here the $x_j$ for $j \in [N]$ represents the sample points on the evenly spaced scan coordinates, and $f_j$ for $j \in [N]$ is the sample on the Fourier space. A discrete Fourier transform can be computed through matrix multiplication with $\mbf{F}$. A similar approach can be done to implement a two-dimensional Fourier transform  by applying the Kronecker product of two one-dimensional Fourier matrices \citep[eq.5.68]{jain1989fundamentals}.

$$
\mathbf{F}_{2D} = \mathbf{F} \otimes \mathbf{F} \in \C^{N^2 \times N^2}.
$$
The Kronecker product constructs a block matrix with the total dimension of the product of the original dimension of two matrices. It should be noted that it is also possible to apply the Kronecker product even if both matrices are not square.

Suppose we have four-dimensional \ac{STEM} data $\mathbf{I}_s \in \R^{N\times N}$ for evenly spaced scanning points $s \in [S]$. Hence, we can vectorize our datasets into $\mathbf{I} \in \R^{S \times N^2}$, where the row and column space represent all scanning points and the dimension of the microscope's detector, respectively. The Fourier transform along the scan coordinates can be done by calculating the matrix product between the two-dimensional Fourier matrix and the data\- sets as expressed in the following

$$
\hat{\mbf{I}} = \mathbf{F}_{2D} \mathbf{I} \in \C^{S \times N^2}.
$$
Let us write the matrix as a collection of all vectors on the column space $\mathbf{F}_{2D} = \left(\mathbf{f}_1,\mathbf{f}_2,\dots,\mathbf{f}_S\right) \in \C^{S \times S}$ and the dataset as $\mathbf{I}^T = \left(\mathbf{i}_1,\mathbf{i}_2,\dots,\mathbf{i}_S \right) \in \R^{N^2 \times S}$. Applying the property of matrix product, which can be expressed as the sum of outer product between column and row element of both matrices \citep[Sect.1.1.14, Algorithm 1.1.8]{golub2013matrix}, we can write the following: 
\begin{equation}
\label{eq:outer_prod}
\mathbf{F}_{2D} \mathbf{I} = \sum_{s=1}^S \mathbf{f}_s \mathbf{i}_s^T \in \C^{S \times N^2},
\end{equation}
This sum is trivial to split into partial sums for parts of the input data $\mathbf{I}$, allowing gradual processing with a live updating partial result. A reformulation of the \ac{WDD} algorithm using this approach will be presented in Section \ref{sect:modif_wdd}. A Jupyter notebook demonstrating this numerically equivalent rearrangement of the reference implementation is available at\\ \url{https://github.com/Ptychography-4-0/LiveWDD}.

\subsection{Pre-computed Wiener filter}
In the conventional \ac{WDD} as presented in Section \ref{Sec2:WDD}, we calculate the pro\-be's Wigner function as the initial parameter for deconvolution. The initial guess of the illuminating probe can be generated from the acquisition settings, such as the semiconvergence angle, to compute the circular aperture in Fourier space. Therefore, the autocorrelation of the initial probe can be pre-computed. The complete algorithm is presented in Algorithm \ref{Algo:PreCompute}. 

\begin{figure*}[htb!]
\centering
\includegraphics[scale=0.7]{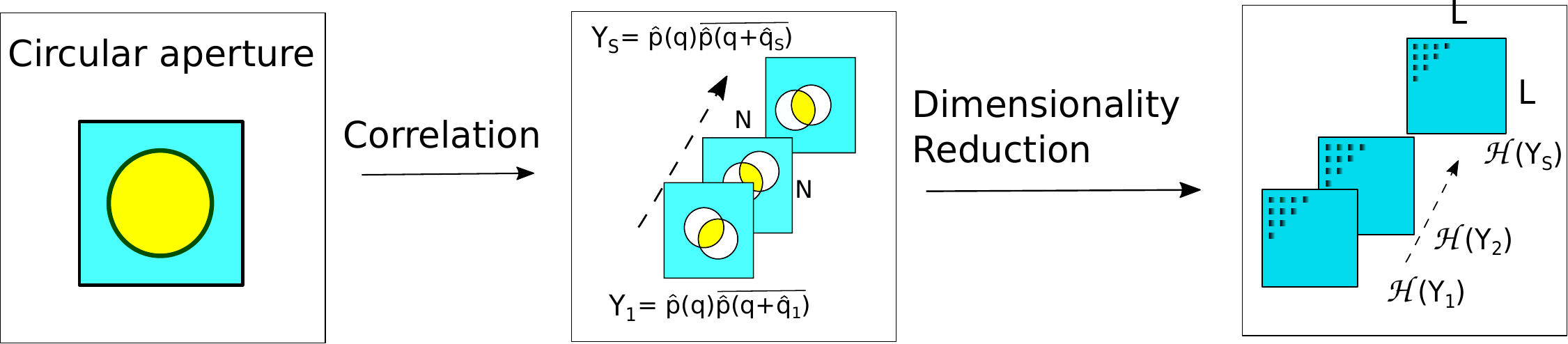}
\caption{\textcolor{black}{Schematic diagram for pre-computed Wiener filter.}}
\label{Fig:schematic_Wiener}
\end{figure*}

\textcolor{black}{The schematic diagram for the pre-computed Wiener filter is presented in Figure \ref{Fig:schematic_Wiener} where starting from the initial probe on the reciprocal space, i.e., circular aperture, we perform the correlation with respect to the shifted position on the spatial frequency. The results present the trotter shape yielding an intersection between both circular apertures. After calculating the correlation function, we continue with the dimensionality reduction to reduce the dimension before using the compressed correlation to calculate the Wiener filter with a small number to avoid zero division $\epsilon$
}.

\begin{algorithm}
\caption{Pre-computed Wiener filter for Live WDD}
\label{Algo:PreCompute}
\begin{algorithmic}[1]
\State \textbf{Initialization:} 
\begin{enumerate}
    \item Initial probe on the Fourier space  $\left(p_{ij}\right)=\hat{p}\left(\mbf q\right)$, for $i,j \in [N]$ i.e., circular apperture, generated from the radius of diffraction patterns.
    \item Choose a small number to avoid zero division $\epsilon$
\end{enumerate}

\For{each  pre-defined index on the spatial frequency coordinate in the set $v \in \{1, 2,\dots,S \}$} 
\State \parbox[t]{\dimexpr\textwidth-\leftmargin-\labelsep-\labelwidth}{Apply transformation to get physical \\coordinates of spatial frequency from acquisition \\setting $\{\hat{\mbf{q}}_1,\hat{\mbf{q}}_2,\dots,\hat{\mbf{q}}_S\}$}
\State \parbox[t]{\dimexpr\textwidth-\leftmargin-\labelsep-\labelwidth}{Construct autocorrelation matrix $\mbf{Y}_v \in \C^{N \times N}$ for \\$v \in [S]$. The element of the matrix is\\
 $
 \left(y_{ij}\right)_v =  \hat{p}\left(\mathbf{q}\right)  \overline{\hat{p}\left(\mathbf{q} + \hat{\mbf{q}}_v \right)} \quad \text{for} \quad i,j \in [N].
 $}
\State \parbox[t]{\dimexpr\textwidth-\leftmargin-\labelsep-\labelwidth}{Calculate dimensionality reduction to the initial\\ probe's autocorrelation $\mcl{H}\left( \mathbf{Y}_v\right)$ for $v \in [S].$}
\State \parbox[t]{\dimexpr\textwidth-\leftmargin-\labelsep-\labelwidth}{Calculate the Wiener filter \\$\mbf{K}_v = \frac{\overline{\mcl{H}\left( \mathbf{Y}_v\right)}}{\card{\mcl{H}\left( \mathbf{Y}_v\right)}^2 + \epsilon} \in \C^{L \times L}$ for $v \in [S].$}
\EndFor

\end{algorithmic}
\end{algorithm} 

\begin{figure}[!htb]
\centering
\includegraphics[scale=0.5]{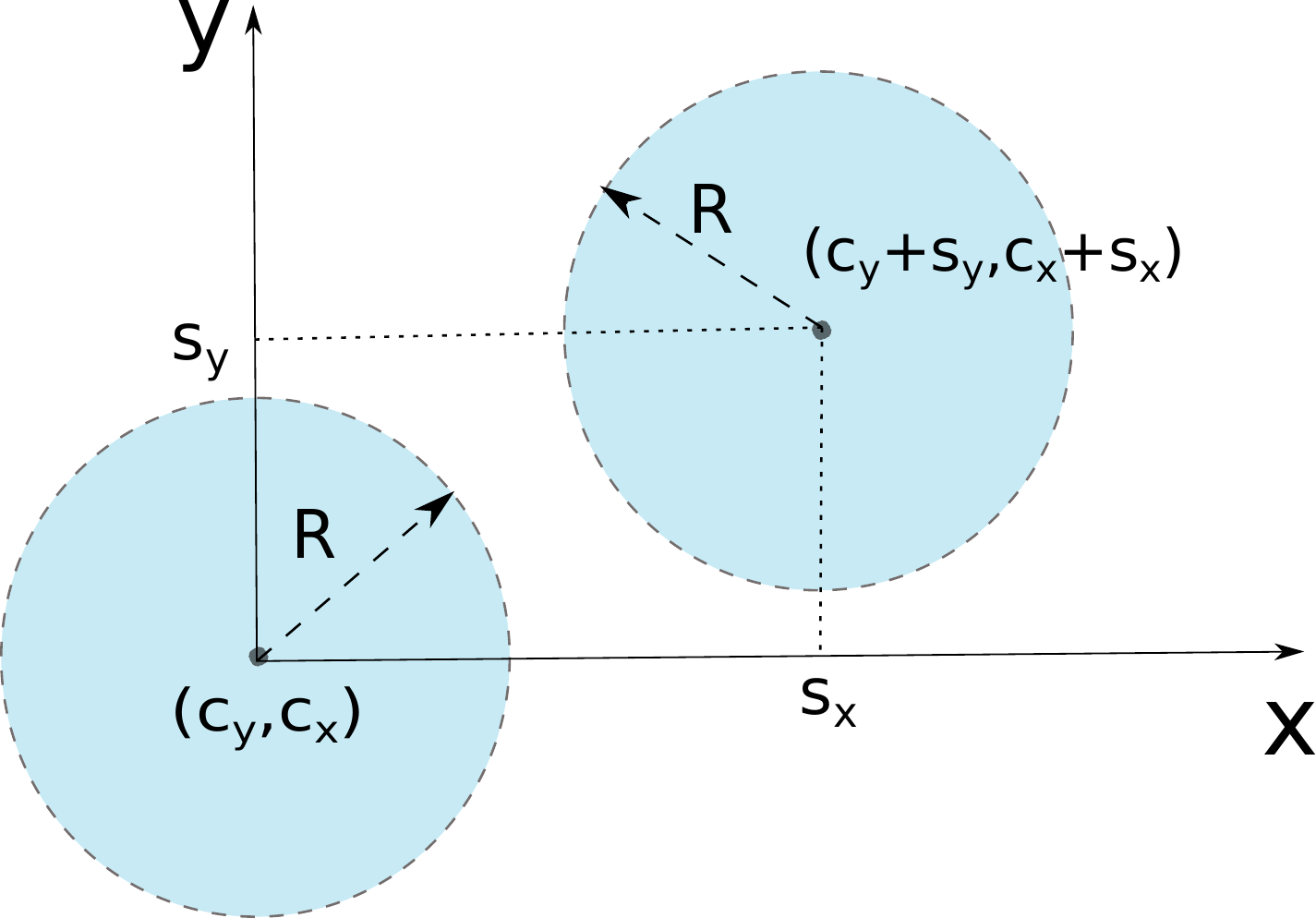}
\caption{Geometry of the autocorrelation of the probe in reciprocal space.}
\label{fig:intersection_circle}
\end{figure}

In typical cases for electron microscopy where the illumination is a focused convergent beam with an angular range limited by a circular aperture, we calculate the autocorrelation between shifted circular apertures. At some spatial frequencies $\hat{\mbf{q}}_v$ we may not have an intersection at all. These frequencies do not contribute to the reconstruction and can be omitted from the calculation. An illustration is given in Figure \ref{fig:intersection_circle}.

The shift depends on the microscope and acquisition parameters, such as electron wavelength $\lambda$, semiconvergence angle $\theta$, the radius of the circle in pixel $R$, and the scanning shift on the real space coordinates for both axes $\Delta_y, \Delta_x$. First of all, let us write the condition when the intersection occurs

\begin{equation}
\begin{aligned}
&s_y^2 +  s_x^2 \leq 4R^2\\
\end{aligned}
\end{equation}
It should be noted that for spatial frequency indexes $v_y, v_x$\newline $\in [S]$, we can write the transformation to physical coordinate given by
$$
s_y = v_y \frac{\lambda R}{\Delta_y S \sin \theta}\quad \text{and} \quad s_x = v_x \frac{\lambda R}{\Delta_x S \sin \theta}
$$
If we have the same scanning shift $\Delta_y = \Delta_x = \Delta$, we can have the condition
\begin{equation}
\begin{aligned}
&\left( v_y^2 +  v_x^2 \right) \frac{\lambda^2}{\Delta^2 S^2 \sin^2 \theta} \leq 4\\
\end{aligned}
\end{equation}
For all combinations of $v_y, v_x \in [S]$, we can find an upper bound $v_y^2 +  v_x^2 \leq 2v^2$, where $v = \text{max}\left(v_y, v_x\right) \in [S]$. Hence, we have
\begin{equation}
\label{eq:low_correlation}
\begin{aligned}
&v \leq \widehat{S} = \frac{\sqrt{2} \Delta \sin \theta }{\lambda}S\\
\end{aligned}
\end{equation}
For specific setting in the acquisition process, we can have $\frac{\sqrt{2} \Delta \sin \theta }{\lambda} < 1$, thereby, only required smaller intersection. \textcolor{black}{To have a concrete number suppose we have a specific value
of measurement settings, i.e., $\theta = 32$ mrad, $\Delta = 0.026$ nm, $\lambda = 2.508$ pm, we have scaling factor $\hat{S} = 0.47S$,
which is smaller than the total spatial frequency index $S$ and can be used to improve the computation time of live processing \ac{WDD}.}

\begin{algorithm}[htb!]
\caption{Modified WDD}
\label{Algo:ModifiedWDD}
\begin{algorithmic}[1]
\State \textbf{Initialization:} 
\begin{enumerate}
    \item Pre-compute Wiener filter $\mbf{K}_v$ for $v \in [S]$ given in Algorithm \ref{Algo:PreCompute} with the number of intersection index $\widehat{S} \leq S$ with applied dimensionality reduction.
    \item Initialize source for sequence of diffraction patterns $\mathbf{I}_s \in \R^{N \times N}$ for each scanning point $s \in [S]$.
    \item Pre-compute DFT matrices $\mbf{F_y} \in \C^{S_y \times S_y}$ and $\mbf{F_x} \in \C^{S_x \times S_x}$ for efficient on-the-fly computation of subsets of $\mathbf{F}_{2D}$.
    \State Allocate and initialize buffer for object function $\mathbf{O} \in \C^{S_y \times S_x}$ with zero.
\end{enumerate}

\For {each given real space scanning point index $s$ 
in the set $\{1, 2,\dots,S \}$} 
\State \parbox[t]{\dimexpr\textwidth-\leftmargin-\labelsep-\labelwidth}{Apply dimensionality reduction $\mathcal{H}\left(\mathbf{I}_s\right) \in \R^{L \times L}$ \\and vectorization to have $\mbf{i}_s =\text{vec}\left(\mathcal{H}\left(\mathbf{I}_s\right)\right) \in  \R^{L^2}$.}
\State \parbox[t]{\dimexpr\textwidth-\leftmargin-\labelsep-\labelwidth}{Calculate subset of the Fourier matrix $\mbf{F}_{2D}$\\ for $s$ from the two pre-computed DFT matrices.}
\State \parbox[t]{\dimexpr\textwidth-\leftmargin-\labelsep-\labelwidth}{Calculate outer product between all column\\ element in the subset of the Fourier matrix $\mbf{F}_{2D}$ \\and vectorized intensity $\mbf{i}_s$ as in \eqref{eq:outer_prod}, $\mathbf{T} = \mbf{f}_s \mbf{i}_s^T \\ \in \C^{S \times L^2}$, where we have $\mbf{t}_v \in \C^{L^2}$ for $v \in [S]$ \\for each row of matrix $\mbf{T}$.\strut}
\For {each  given spatial frequency index with non- \,\,zero intersection $v$ in the set $\{1, 2,\dots, \widehat{S} \}$}
\State \parbox[t]{\dimexpr\textwidth-\leftmargin-\labelsep-\labelwidth}{Reshape the result for each row, \\ i.e., $\text{mat}\left(\mbf{t}_v\right) \in \C^{L \times L}$.}

\State \parbox[t]{\dimexpr\textwidth-\leftmargin-\labelsep-\labelwidth}{Deconvolution process on compressed space \\$\mbf{D}_v = \text{mat}\left(\mbf{t}_v\right) \circ \mbf{K}_v \in \C^{L \times L}$ for $v \in [\widehat{S}]$.}
\State\parbox[t]{\dimexpr\textwidth-\leftmargin-\labelsep-\labelwidth}{Calculate zero frequency by summing all \\elements of matrix $\mathbf{D}_v = \left(d_{lk}\right)_v$ for $l,k \in [L]$. \\ Here we have scalar for each spatial frequency \\$o_v = {\sum_{l=1}^L \sum_{k=1}^L \left(d_{lk}\right)_v}$\strut}
\State\parbox[t]{\dimexpr\textwidth-\leftmargin-\labelsep-\labelwidth}{Replace the element of the non-zero index in\\ spatial frequency with $o_{v}$ \strut}
\EndFor
\State \parbox[t]{\dimexpr\textwidth-\leftmargin-\labelsep-\labelwidth}{We have an update vector\\ $\mbf{o} = \left(o_1,o_2,\dots, o_S\right) \in \C^S$. This is added to the \\buffer allocation matrix for each given scanning \\point
$
\mathbf{O} := \mathbf{O} + \text{mat}\left(\mbf{o}\right) \in \C^{S_y \times S_x}
$}
\EndFor
\State \parbox[t]{\dimexpr\textwidth-\leftmargin-\labelsep-\labelwidth}{Applying inverse Fourier transform and complex\\ conjugate on set of spatial frequencies to get real \\space coordinate of the specimen transfer function \\$\overline{\mathcal{F}^{-1} \left( \mathbf{O}\right)}$ }
\end{algorithmic}
\end{algorithm} 

\subsection{Modified \ac{WDD}}\label{sect:modif_wdd}
\begin{figure*}[htb!]
\centering
\includegraphics[scale=1.]{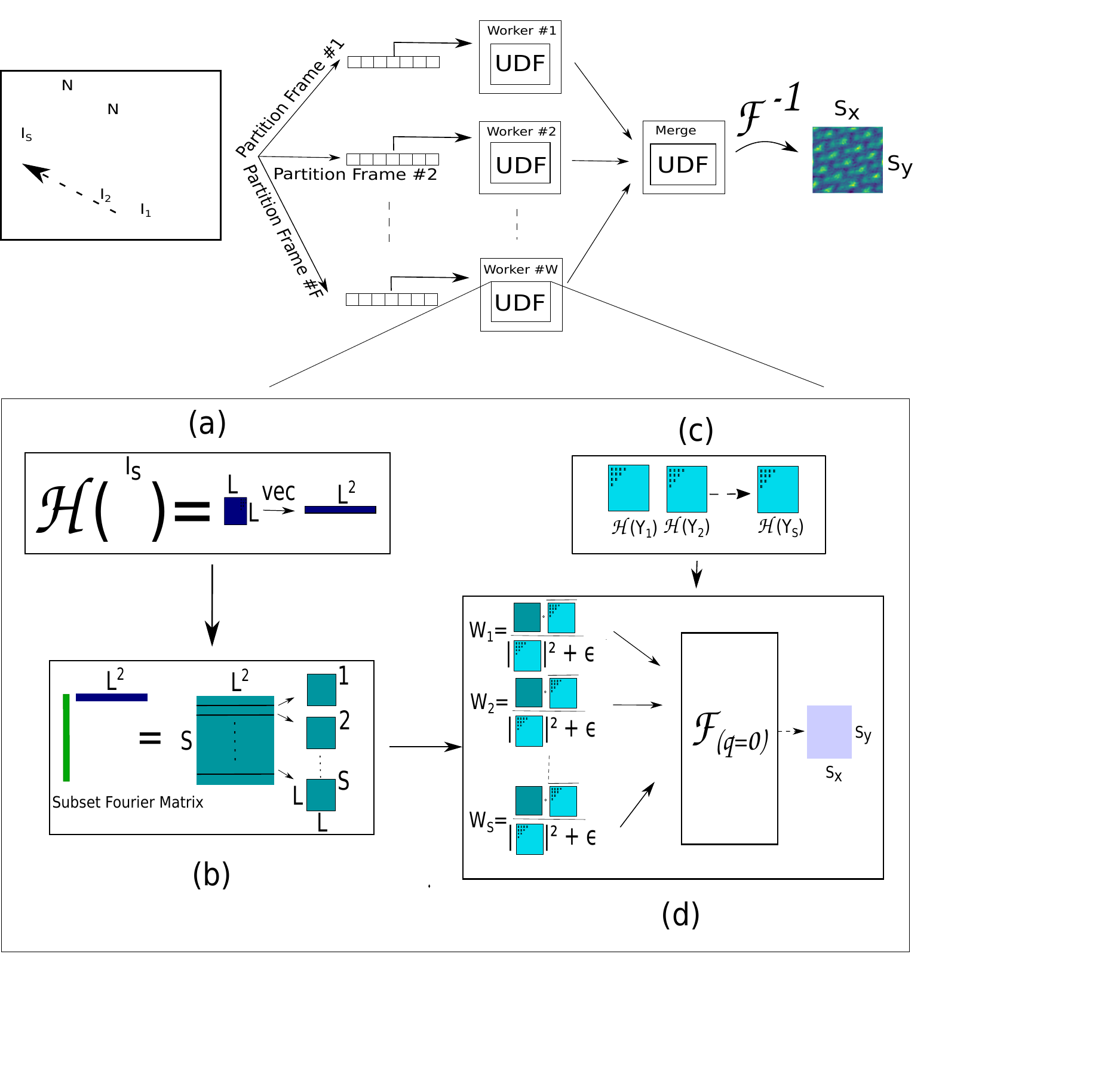}
\vspace*{-2.5cm}
\caption{\textcolor{black}{The schematic diagram for running the Live \ac{WDD} as a \ac{UDF} in LiberTEM-live. The data stream from the detector is split into partitions, which are streamed to worker processes via queues. Each diffraction pattern is processed in the \ac{UDF} as follows: (a) Applying dimensionality reduction and vectorization, (b) Processing the compressed frame with a subset of the Fourier matrix, (c) Accessing the pre-computed Wiener filter to perform a single reconstruction as presented in (d). The merge function sums up the contributions from each partition, and the final reconstruction in real space is determined by the two-dimensional inverse Fourier transform.
}}
\label{Fig:schematic_LWDD}
\end{figure*}

Combining dimensionality reduction, gradual processing, as well as the pre-computed Wiener filter, we present the modified WDD for live reconstruction in Algorithm \ref{Algo:ModifiedWDD}. Sin\-ce we compress and process the diffraction patterns per scanning point in Algorithm \ref{Algo:ModifiedWDD}, the total reconstruction of the specimen transfer function is updated gradually, starting from a zero-initialized matrix. Furthermore, the updates from subsets of the input data can be computed independently, which allows trivial parallelization. This algorithm for Live \ac{WDD} is implemented as a \ac{UDF} for LiberTEM-live \citep{clausen2020libertem}. The complete schematic diagram for an implementation of \ac{WDD} as a LiberTEM \ac{UDF} is presented in Figure \ref{Fig:schematic_LWDD}. It is available as Open Source at \url{https://github.com/Ptychography-4-0/LiveWDD}.

\section{Time and Space Complexity \label{Sec:Complexity}}
Here we discuss the analysis of the Live \ac{WDD} algorithm and compare it with conventional \ac{WDD}. For both approaches, we derive the time and space complexity required to perform the steps specified by the algorithm. In this section, we denote the total scanning points as $S$ for raster position on both $x,y$ axis, i.e., $S = S_x \times S_y$. In addition, the dimension of the detector is denoted as $N \times N$.  In Table \ref{tab:time_space_comp} we provide a summary for the time and space complexity analysis for both algorithms. 
\begin{table}[htb!]
\centering
\caption{Computational complexity of conventional \ac{WDD} and  Live \ac{WDD}}
\label{tab:time_space_comp}
\footnotesize
\begin{tabular}{ |c|c|c|c| } 
\hline
Complexity & Conventional \ac{WDD} & Live \ac{WDD} \\
\hline
Time & $\mcl{O}\left(SN^2 \log SN^2\right)$ & $\text{max}\left( \mcl{O}(SN^2L) , \mcl{O}(S^2L^2)\right)$ \\ 
\hline
Space & $\mcl{O}\left(SN^2\right)$ &  $\mcl{O}({S}L^2)$ \\ 
\hline
\end{tabular}
\vspace*{-0.5cm}
\end{table}

For space complexity it appears that the Live \ac{WDD} scales better compared to conventional \ac{WDD} since $L \ll N$. However, for time complexity it highly depends on the total scanning points $S$ and the logarithmic factor $\log(SN^2)$ compared to the low dimension $L$, here we use $L = 16$.  The derivation of complexity analysis is provided in Section \ref{Sec:Complexity_ConvWDD} and Section \ref{Sec:Complexity_LiveWDD}, respectively.
\subsection{Conventional \ac{WDD}}\label{Sec:Complexity_ConvWDD}
First of all, conventional \ac{WDD} performs a Fourier transform along the scanning position in real space to obtain the spatial frequencies. The computational complexity of a fast Fourier transform for all scanning points is $\mcl{O}(SN^2\\ \log S)$.
For each spatial frequency, we calculate the autocorrelation of the probe, which gives us computation time $\mcl{O}\left(N^2\right)$. Afterward, we calculate the Wigner distribution function by applying inverse Fourier transform on the autocorrelation as well as the intensity of diffraction patterns, for each taking $\mcl{O}\left(N^2 \log N^2\right)$. The process is followed by applying Wiener filtering or deconvolution that requires $\mcl{O}\left(N^2 \right)$. Since we have to calculate all spatial frequencies, the computation is therefore $S \left( \mcl{O}(N^2\right)+ \mcl{O}(N^2 \log N^2))$, which gives total for processing all spatial frequencies $\mcl{O} \left(SN^2 \log N^2 \right)$. The next step is to calculate the Fourier transform of all deconvolution data before taking only zero reciprocal space, i.e., $\mbf{q} = 0$, hence we perform operation $\mcl{O} \left(SN^2 \log N^2 \right)$. In the last step, we apply an inverse Fourier transform for estimated object which requires $\mcl{O}\left( S \log S\right)$. The total time computation for conventional \ac{WDD} then $\mcl{O}\left(SN^2 \log(S)\right) + \mcl{O} \left(SN^2 \log N^2 \right) + \mcl{O}\left( S \log S\right)$. Simplification gives us time complexity of conventional \ac{WDD} as $\mcl{O}\left(SN^2 \log SN^2\right)$.

For space complexity, we start with total memory allocation to store all four-dimensional datasets to perform a Fourier transform on scanning points on the real space, i.e., to obtain the spatial frequencies, which requires \\$\mcl{O}\left(SN^2\right)$. For each spatial frequency, we have to process the data with dimension $\mcl{O}\left(N^2\right)$. In the last process to calculate the estimated object we have space complexity $\mcl{O}\left(S\right)$. Thereby, total space complexity of conventional \ac{WDD} is $\mcl{O}\left(SN^2\right) + \mcl{O}\left(N^2\right) + \mcl{O}\left(S\right)$, which then we have scaling for space complexity $\mcl{O}\left(SN^2\right)$.

\subsection{Live \ac{WDD}}\label{Sec:Complexity_LiveWDD}
The Live WDD presented here includes a compression step where the dimension of the compressed data is $L \times L $, much smaller than detector size ($L \ll N$). For the time complexity of Live \ac{WDD}, we separate the processing into the computation of the Wiener filter in Algorithm \ref{Algo:PreCompute} and the actual data processing.
The Wiener filter requires $\mcl{O}(N^2)$ for computing the autocorrelation function. Since we apply a dimensionality reduction, which is performed with three matrix multiplications, we have $\mcl{O}(LN^2 + L^2N)$. At the end of the process, we compute the element-wise division for the Wiener filter with complexity $\mcl{O}(L^2)$ and incorporate the process for all spatial frequencies $S$. Therefore,  we have time complexity for the pre-computed Wiener filter as $\mcl{O}({S}N^2L)$. 
For processing each scanning point, we calculate the dimensionality reduction for each diffraction pattern with complexity $\mcl{O}(L^2N  + N^2 L)$. Afterward, the computation of the outer product between each column of Fourier matrix with a compressed diffraction pattern requires $\mcl{O}\left(S L^2\right)$. The next process is the deconvolution with the pre-compu\-ted Wiener filter for all spatial frequencies which takes $\mcl{O}\left(SL^2\right)$. Getting zero frequencies and  the update data have complexity $\mcl{O}\left(SL^2\right)$ and $\mcl{O}\left(S\right)$, respectively. Thereby, processing with all real space scanning points is given by $\mcl{O}(S L^2N  + S N^2L) +  \mcl{O}(S^2L^2) + \mcl{O}(S^2)$. In the last step, to get the estimated object on the real space with dimension $S$, we apply an inverse Fourier transform that requires $\mcl{O}\left(S \log S\right)$. 

In the end, for Live \ac{WDD} we have total time complexity $\mcl{O}\left(S L^2N  + S N^2L\right) + \mcl{O}(S^2L^2 ) +  \mcl{O}(S^2 ) +  \mcl{O}\left(S \log S\right) $, which then gives us $\text{max}\left( \mcl{O}(SN^2L) , \mcl{O}(S^2L^2)\right)$. The result highly depends on the dimension of $N$ and ${S}$, where for a larger detector dimension than total spatial frequency, we have time complexity $\mcl{O}\left(S N^2L\right)$. On the contrary, if we have a large field of view, there is a possibility then we have time complexity $\mcl{O}\left(S^2L^2\right)$. 

Regarding the space complexity, in Live \ac{WDD} we can process the data per frame, i.e. per diffraction pattern, the size of which corresponds to the detector's dimension. In comparison, conventional \ac{WDD} stores the complete four-dimensional dataset to compute the spatial frequencies. In addition, we apply a dimensional reduction technique that only requires $\mcl{O}\left(L^2 \right)$. For each scanning point on the real space as well as the non-zero intersection on the spatial frequency, the algorithm only requires $\mcl{O}(L^2)$ except for computing the estimated object that has dimension $\mcl{O}(S)$. For the pre-computed Wiener filter, we have to store $\widehat{S} \leq S$ non-zero intersection spatial frequencies for the autocorrelation process with compressed dimension $L \times L$. Hence we have $\mcl{O}({S}L^2)$ space complexity of the pre-computed Wiener filter. In total we have $\mcl{O}({S}L^2) + \mcl{O}( S ) + \mcl{O}(L^2)$ and give us complexity $\mcl{O}({S}L^2)$, which still better than conventional \ac{WDD}.

In Section \ref{Sec:Numeric_Results}, we compare real-world time and space use of the proposed Live \ac{WDD} and conventional \ac{WDD} on various datasets.

\section{Simulation and Experimental Datasets}
\begin{figure}[!htb]
\centering
\includegraphics[scale=0.4]{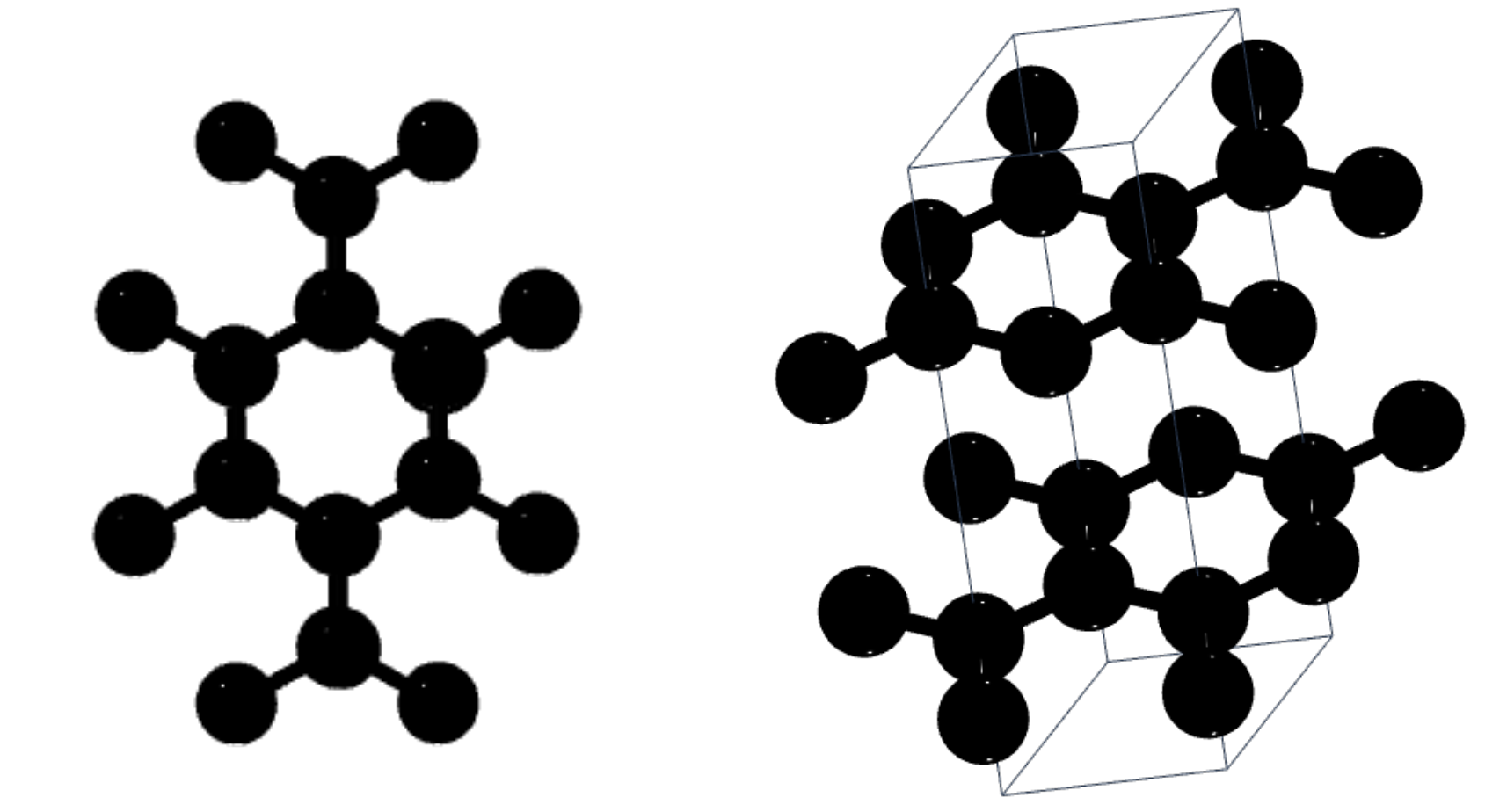}
\caption{Structure of graphene. The structure is downloaded from $\texttt{materialsproject.org}$\citep{jain2013commentary}}
\label{fig: datasets_pytycho}
\end{figure}

The information of datasets and parameter settings to evaluate Live \ac{WDD} are given in \citep{t_pennycook_2021_4476506}  for simulated graphene. The specimen has a hexagonal lattice structure, shown in Figure \ref{fig: datasets_pytycho}. The four-dimensional \ac{STEM} data are generated by parameter settings presented in Table \ref{tab:params}.
 
\begin{table}[!htb]
  \caption{Parameters for simulated graphene \citep{t_pennycook_2021_4476506}}
\label{tab:params}
 \centering
  \begin{tabular}{lll}
    \toprule
     \bfseries Parameters & \bfseries Graphene \\
     \bottomrule
      Rotation (deg) & $0.0$ \\
      Semiconv. angle (mrad) & $30$ \\
      Accel. voltage (keV) & $60$ \\
      Scanning step size (nm) & $0.02$ \\
      Scanning points (Sy, Sx) & $\left(64,65\right)$ \\
      Detector size (pixel)& $\left( 256,256\right)$ \\
    \bottomrule
  \end{tabular} 
\end{table}
We can also add the effect of Poissonian noise to the simulated diffraction patterns data. Suppose we have dose level per pixel represented by variable $\nu$ that has a unit $\,e^{-}/$\AA$^2$ . The model used to generate a noisy dataset for each intensity of diffraction patterns $\mbf{I}_s \in \R^{N \times N}$ for $s \in [S]$ is given as follows

$$
\texttt{Poisson}\left(\nu \mbf{\widetilde{I}}_s\right) \in \R^{N \times N},
$$
where the $\mbf{\widetilde{I}}_s$ is normalisation of the diffraction pattern for each scanning point  and $\texttt{Poisson}$ is a function to generate Poisson distribution applied to our dataset. It should be noted that this function preserves the dimension of the data.

\begin{figure}[!htb]
\centering
\includegraphics[scale=0.7]{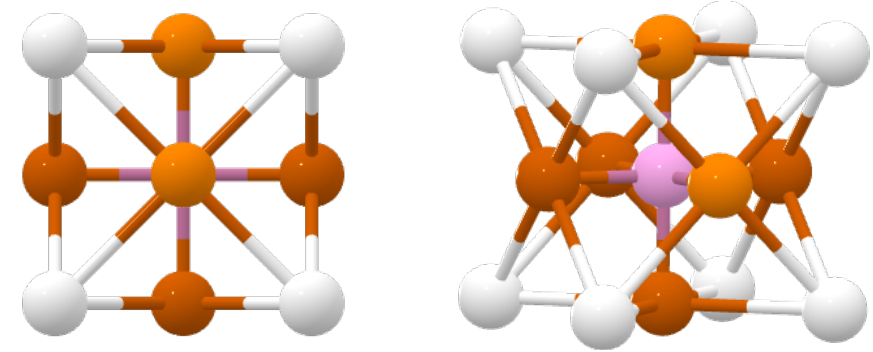}
\caption{Structure of \ac{SrTiO$_3$} where Strontium, Titanium, and Oxygen are visualized as white, purple, and red colors. The structure is downloaded from $\texttt{materialsproject.org}$\citep{jain2013commentary}}
\label{fig: datasets_pytycho_exp}
\end{figure}

Additionally, we also used an experimental datasets of \ac{SrTiO$_3$} specimen acquired using Medipix Merlin EM detector \citep{strauch2021high}, where the structure is presented in Figure \ref{fig: datasets_pytycho_exp}. The parameters setting of this dataset is presented in Table \ref{tab:params_exp}. Complete information about this dataset can be directly seen \citep{strauch2021high}.
\begin{table}[!htb]
  \caption{Parameters for experimental dataset \ac{SrTiO$_3$} \citep{strauch2021high}}
\label{tab:params_exp}
 \centering
  \begin{tabular}{llll}
    \toprule
     \bfseries Parameters & \bfseries \ac{SrTiO$_3$} \\
     \bottomrule
      Rotation (deg) & $88$  \\
      Semiconv. angle (mrad) & $22.13$\\
      Accel. voltage (keV) & $300$ \\
      Scanning step size (nm) & $0.0127$ \\
      Scanning points (Sy, Sx) & $\left(128,128\right)$ \\
      Detector size (pixel)& $\left( 256,256\right)$\\
    \bottomrule
  \end{tabular}
\vspace{-0.4cm}
\end{table}

\section{Numerical Results}\label{Sec:Numeric_Results}
We present numerical evaluations of the Live \ac{WDD} in terms of reconstruction, computation time, and memory allocation. For Live \ac{WDD}, we also present how the live reconstruction evolves from partial results. As a comparison to the existing \ac{WDD} implementation, we refer to the implementation in \citep{yang2016simultaneous} as the reference to check the dynamic range of the phase, where the source code is available in  
\url{ https://gitlab.com/ptychoSTEM/ptychoSTEM } and \url{https://gitlab.com/PyPtychoSTEM/PyPtychoSTEM}. We use the same reconstruction parameters in \ac{WDD} as given in the source code, for instance, the small constant $\epsilon$ for the Wiener filter.

The evaluation is performed individually on the same workstation with AMD EPYC 7543P with $32$ CPU with $64$ threads operating at a base frequency of $2.80$\, GHz, and $512$\, GB DDR4 RAM with an operating frequency of $3.2$\, GHz. It should be noted that the simulation for each algorithm is performed without a noise background, i.e., no other processes or algorithms were running during the evaluation.

\subsection{Reconstruction}
In this section, we performed numerical comparisons of the Live \ac{WDD} and the conventional \ac{WDD} given in the PyPtychoSTEM, with a similar setting for both algorithms, e.g., $\epsilon = 0.01$. In the first part, we focus on the evaluation of noise-free conditions for graphene datasets. The second part covers the performance of algorithms when applied to data that is degraded by Poissonian noise corresponding to different dose levels. Since the goal is to show the specimen transfer functions, we present the final phase reconstruction from the Live \ac{WDD}.

\begin{figure}[!htb]
\centering
     \scalebox{0.4}{
\pgfplotsset{every tick label/.append style={font=\Huge}}
\begin{tikzpicture}
\begin{groupplot}[group style={group size=2 by 1 ,horizontal sep=10em},width = 0.5*\textwidth, height=(0.5)*\textwidth]
\nextgroupplot[
colorbar,
colorbar style={ylabel={},scaled y ticks=false,yticklabel style={/pgf/number format/fixed}},
colormap/blackwhite,
hide x axis,
hide y axis,
point meta max=0.0817124843597412,
point meta min=-0.0385599210858345,
tick align=outside,
tick pos=left,
title={\Huge (a)},
x grid style={white!69.0196078431373!black},
xmin=-0.5, xmax=64.5,
xtick style={color=black},
y dir=reverse,
y grid style={white!69.0196078431373!black},
ymin=-0.5, ymax=63.5,
ytick style={color=black}
]
\addplot graphics [includegraphics cmd=\pgfimage,xmin=-0.5, xmax=64.5, ymin=63.5, ymax=-0.5] {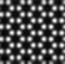};

\nextgroupplot[
colorbar,
colorbar style={ylabel={},scaled y ticks=false,yticklabel style={/pgf/number format/fixed}},
colormap/blackwhite,
hide x axis,
hide y axis,
point meta max=0.089402936398983,
point meta min=-0.051781639456749,
tick align=outside,
tick pos=left,
title={\Huge (b)},
x grid style={white!69.0196078431373!black},
xmin=-0.5, xmax=64.5,
xtick style={color=black},
y dir=reverse,
y grid style={white!69.0196078431373!black},
ymin=-0.5, ymax=63.5,
ytick style={color=black},
]
\addplot graphics [includegraphics cmd=\pgfimage,xmin=-0.5, xmax=64.5, ymin=63.5, ymax=-0.5] {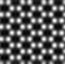};
\end{groupplot}
\end{tikzpicture}}  

\caption{Phase reconstruction of simulated graphene with (a). PyPtychoSTEM and (b). Live Processing \ac{WDD}}
\label{fig:phase_graphene} 
\end{figure}

\begin{figure*}[!htb]
\centering
     \scalebox{0.4}{
\pgfplotsset{every tick label/.append style={font=\Huge}}
\begin{tikzpicture}

\begin{groupplot}[group style={group size=4 by 2,horizontal sep=9em},width = 0.5*\textwidth, height=(0.5)*\textwidth]
\nextgroupplot[
colorbar,
colorbar style={ylabel={},scaled y ticks=false,yticklabel style={/pgf/number format/fixed}},
colormap/blackwhite,
hide x axis,
ylabel = \Huge PyPtychostem,
ticks=none,
point meta max=0.93514041185379,
point meta min=-0.807466387748718,
tick align=outside,
tick pos=left,
title={\Huge (a)},
x grid style={white!69.0196078431373!black},
xmin=-0.5, xmax=64.5,
xtick style={color=black},
y dir=reverse,
y grid style={white!69.0196078431373!black},
ymin=-0.5, ymax=63.5,
ytick style={color=black}
]
\addplot graphics [includegraphics cmd=\pgfimage,xmin=-0.5, xmax=64.5, ymin=63.5, ymax=-0.5] {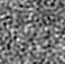};

\nextgroupplot[
colorbar,
colorbar style={ylabel={},scaled y ticks=false,yticklabel style={/pgf/number format/fixed}},
colormap/blackwhite,
hide x axis,
hide y axis,
point meta max=0.354699492454529,
point meta min=-0.298538684844971,
tick align=outside,
tick pos=left,
title={\Huge (b)},
x grid style={white!69.0196078431373!black},
xmin=-0.5, xmax=64.5,
xtick style={color=black},
xticklabels={},
y dir=reverse,
y grid style={white!69.0196078431373!black},
ymin=-0.5, ymax=63.5,
ytick style={color=black},
yticklabels={}
]
\addplot graphics [includegraphics cmd=\pgfimage,xmin=-0.5, xmax=64.5, ymin=63.5, ymax=-0.5] {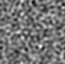};

\nextgroupplot[
colorbar,
colorbar style={ylabel={},scaled y ticks=false,yticklabel style={/pgf/number format/fixed}},
colormap/blackwhite,
hide x axis,
hide y axis,
point meta max=0.135960400104523,
point meta min=-0.100082859396935,
scaled x ticks=manual:{}{\pgfmathparse{#1}},
scaled y ticks=manual:{}{\pgfmathparse{#1}},
tick align=outside,
tick pos=left,
title={\Huge (c)},
x grid style={white!69.0196078431373!black},
xmin=-0.5, xmax=64.5,
xtick style={color=black},
xticklabels={},
y dir=reverse,
y grid style={white!69.0196078431373!black},
ymin=-0.5, ymax=63.5,
ytick style={color=black},
yticklabels={}
]
\addplot graphics [includegraphics cmd=\pgfimage,xmin=-0.5, xmax=64.5, ymin=63.5, ymax=-0.5] {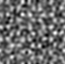};

\nextgroupplot[
colorbar,
colorbar style={ylabel={},scaled y ticks=false,yticklabel style={/pgf/number format/fixed}},
colormap/blackwhite,
hide x axis,
hide y axis,
point meta max=0.101577572524548,
point meta min=-0.0563766621053219,
scaled x ticks=manual:{}{\pgfmathparse{#1}},
scaled y ticks=manual:{}{\pgfmathparse{#1}},
tick align=outside,
tick pos=left,
title={\Huge (d)},
x grid style={white!69.0196078431373!black},
xmin=-0.5, xmax=64.5,
xtick style={color=black},
xticklabels={},
y dir=reverse,
y grid style={white!69.0196078431373!black},
ymin=-0.5, ymax=63.5,
ytick style={color=black},
yticklabels={}
]
\addplot graphics [includegraphics cmd=\pgfimage,xmin=-0.5, xmax=64.5, ymin=63.5, ymax=-0.5] {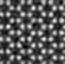};

\nextgroupplot[
colorbar,
colorbar style={ylabel={},scaled y ticks=false,yticklabel style={/pgf/number format/fixed}},
colormap/blackwhite,
ylabel = \Huge Live \ac{WDD},
ticks=none,
hide x axis,
point meta max=0.872675657272339,
point meta min=-0.863502979278564,
tick align=outside,
tick pos=left,
x grid style={white!69.0196078431373!black},
xmin=-0.5, xmax=64.5,
xtick style={color=black},
y dir=reverse,
y grid style={white!69.0196078431373!black},
ymin=-0.5, ymax=63.5,
ytick style={color=black}
]
\addplot graphics [includegraphics cmd=\pgfimage,xmin=-0.5, xmax=64.5, ymin=63.5, ymax=-0.5] {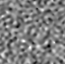};

\nextgroupplot[
colorbar,
colorbar style={ylabel={},scaled y ticks=false,yticklabel style={/pgf/number format/fixed}},
colormap/blackwhite,
hide x axis,
hide y axis,
point meta max=0.279761433601379,
point meta min=-0.218432053923607,
scaled y ticks=manual:{}{\pgfmathparse{#1}},
tick align=outside,
tick pos=left,
x grid style={white!69.0196078431373!black},
xmin=-0.5, xmax=64.5,
xtick style={color=black},
y dir=reverse,
y grid style={white!69.0196078431373!black},
ymin=-0.5, ymax=63.5,
ytick style={color=black},
yticklabels={}
]
\addplot graphics [includegraphics cmd=\pgfimage,xmin=-0.5, xmax=64.5, ymin=63.5, ymax=-0.5] {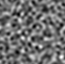};

\nextgroupplot[
colorbar,
colorbar style={ylabel={},scaled y ticks=false,yticklabel style={/pgf/number format/fixed}},
colormap/blackwhite,
hide x axis,
hide y axis,
point meta max=0.135389238595963,
point meta min=-0.0934751555323601,
scaled y ticks=manual:{}{\pgfmathparse{#1}},
tick align=outside,
tick pos=left,
x grid style={white!69.0196078431373!black},
xmin=-0.5, xmax=64.5,
xtick style={color=black},
y dir=reverse,
y grid style={white!69.0196078431373!black},
ymin=-0.5, ymax=63.5,
ytick style={color=black},
yticklabels={}
]
\addplot graphics [includegraphics cmd=\pgfimage,xmin=-0.5, xmax=64.5, ymin=63.5, ymax=-0.5] {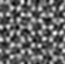};

\nextgroupplot[
colorbar,
colorbar style={ylabel={},scaled y ticks=false,yticklabel style={/pgf/number format/fixed}},
colormap/blackwhite,
hide x axis,
hide y axis,
point meta max=0.104609042406082,
point meta min=-0.0636264756321907,
scaled y ticks=manual:{}{\pgfmathparse{#1}},
tick align=outside,
tick pos=left,
x grid style={white!69.0196078431373!black},
xmin=-0.5, xmax=64.5,
xtick style={color=black},
y dir=reverse,
y grid style={white!69.0196078431373!black},
ymin=-0.5, ymax=63.5,
ytick style={color=black},
yticklabels={}
]
\addplot graphics [includegraphics cmd=\pgfimage,xmin=-0.5, xmax=64.5, ymin=63.5, ymax=-0.5] {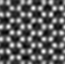};
\end{groupplot}

\end{tikzpicture}}  

\caption{Phase reconstruction of simulated graphene with different dose levels (a) $10^2\,e^{-}/$\AA$^2$, (b) $10^3\,e^{-}/$\AA$^2$,(c) $10^4\,e^{-}/$\AA$^2$, (d) $10^5\,e^{-}/$\AA$^2$}
\label{fig:phase_graphene_dose} 
\end{figure*}

Figure \ref{fig:phase_graphene}  shows the reconstruction of the graphene dataset for both the PyPtychoSTEM implementation of \ac{WDD} and the Live \ac{WDD} for noise-free conditions. It can be seen that the Live \ac{WDD} can reconstruct the specimen transfer function with the correct orientation of the atom represented by the phase and similar dynamic range. In the next evaluation, we conduct numerical evaluation for different dose levels, namely for $\nu \in \{ 10^{2},10^{3},10^{4},10^{5}\}$ \\$\,e^{-}/$\AA$^2$. The reconstruction for noisy setting is depicted in Figure \ref{fig:phase_graphene_dose}.

Compared to the conventional \ac{WDD}, which performs faithful reconstruction starting from dose $\nu = 10^5\,e^{-}/$\AA$^2$, Live \ac{WDD} yields a recognizable reconstruction starting from $\nu = 10^4\,e^{-}/$\AA$^2$. Hence, it can be seen that the Live \ac{WDD} is more robust against Poissonian noise. Besides the property of flexible dimensional reduction, the arbitrary order of the Hermite-Gauss functions can be seen as a low-pass filter, where the width $\sigma$ is optimized so that the any pixel noise is strongly suppressed by the dimensionality reduction, as in Figure \ref{fig:hermite_gauss_function}.

\begin{figure*}[!htb]
\centering
     \scalebox{0.4}{\input{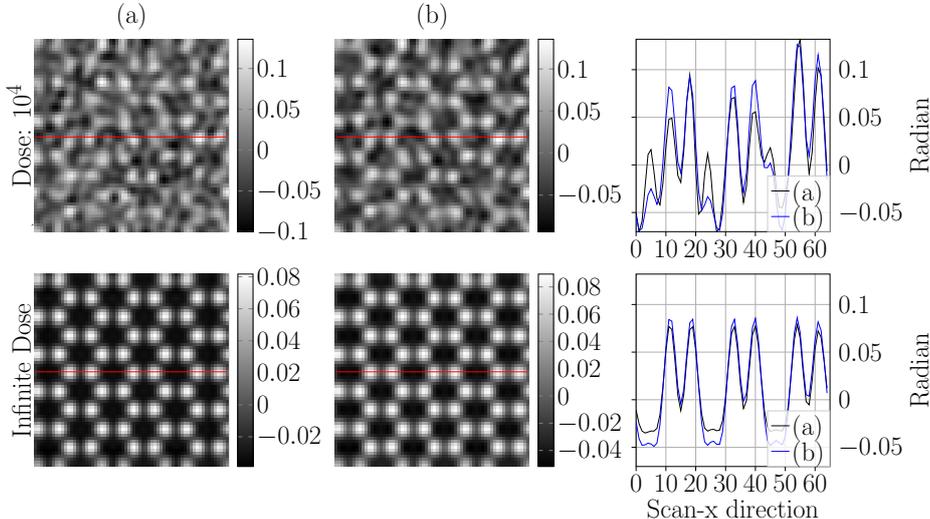}}  

\caption{Phase reconstruction of simulated graphene with dose level $10^4\,e^{-}/$\AA$^2$ and infinite dose for both algorithms (a). PyPtychoSTEM and (b). Live WDD. The line scan reconstruction is also presented where the scan location on Y-direction: $32$ is shown with a red line.}
\label{fig:phase_line_scan} 
\end{figure*}

A line scan through the reconstruction at dose level $10^4\,e^{-}/$\AA$^2$ is presented in Figure \ref{fig:phase_line_scan}, which shows that the noise affects the phase reconstruction of both algorithms. It can be seen that the Live \ac{WDD} has better noise suppression than PyPtychoSTEM, where the latter requires more dose to reliably find atom positions in the phase reconstruction. For infinite dose, both algorithms present the same atom positions with a different value range for Live \ac{WDD} compared to PyPtychoSTEM. A difference is to be expected because Live \ac{WDD} reduces the dimensionality using Hermite-Gauss functions, while PyPtychoSTEM uses Fou\-rier space without dimensionality reduction. That means the two methods are not numerically equivalent.

\subsection{Computation time}

The numerical computation time for Live \ac{WDD} compared to the conventional \ac{WDD} given in \citep{yang2016simultaneous}  is discussed. The evaluation is presented in Table \ref{tab:numeric_inc_det} and Table \ref{tab:numeric_inc_scan}, where we measure the median, as well as the standar deviation of computation time for live and conventional \ac{WDD} for different dimension of datasets generated from Graphene in \citep{t_pennycook_2021_4476506}. 
\begin{table}[htpb]
\centering
\caption{Numerical median computation time in seconds for PyPtychoSTEM and Live \ac{WDD} for fixed dimension scanning points $S_y = S_x = 128$ and increasing  detector $N_y = N_x$.}
\label{tab:numeric_inc_det}
\small
\begin{tabular*}{0.45\textwidth}{Y @{\extracolsep{\fill}} YYY}
\hline
    \multicolumn{1}{c}{Dimension} & \multicolumn{1}{c }{PyPtychoSTEM} & \multicolumn{1}{c}{Live WDD} \\\midrule
     128 &   48.76 \pm 0.14  &  1.28 \pm 0.07  \\
     256 &  145.51 \pm 2.16  &  2.21 \pm 0.069 \\
     512 &  631.71 \pm 9.05  & 12.60 \pm 0.65  \\
    1024 &   -              & 37.42 \pm 0.85  \\
    \bottomrule
\end{tabular*}
\end{table}
To investigate the effect of both increasing scanning points and detector dimensions on the computation time, we also observe both parameter settings, where we use the convention $S_y, S_x$ as the number of scanning points for both axes in the raster scan. Dimension of detector is given by $N_y, N_x$. 

We perform $10$ trials to measure the numerical computation time for both algorithms. From these measurements we show the median computation time. It can be seen that PyPtychoSTEM requires more memory than available to accomplish the reconstruction. As discussed in the Section \ref{Sec:Complexity}, accommodating an entire dataset requires a large memory allocation for PyPtychoSTEM and impinges on the computation performance in general. 
\begin{table}[htpb]
\centering
\caption{Numerical median computation time in seconds for PyPtychoSTEM and Live \ac{WDD} for fixed dimension detector $N_y = N_x = 128$ and increasing scanning points $S_y = S_x$.}
\label{tab:numeric_inc_scan}
\small
\begin{tabular*}{0.45\textwidth}{Y @{\extracolsep{\fill}} YYY}
\hline
    \multicolumn{1}{c}{Dimension} & \multicolumn{1}{c }{PyPtychoSTEM} & \multicolumn{1}{c}{Live WDD} \\\midrule
     128 &   48.76 \pm 0.14  &  1.28 \pm 0.07  \\
     256 &  199.12 \pm 0.60  &  14.82 \pm 0.05 \\
     512 &  824.04 \pm 7.45  & 228.59 \pm 0.18  \\
    1024 &   -              & 3048.48 \pm 5.94  \\
    \bottomrule
\end{tabular*}
\end{table}
In all cases, Live \ac{WDD} performs faster numerical computation than conventional \ac{WDD} implemented in PyPtychoSTEM. However, when the same dimension for both scanning points and detector size is evaluated, i.e., $(1024,\\1024)$, we observe the computation time increases approximately a hundred-fold due to the quadratic scaling in the time complexity, as discussed in Section \ref{Sec:Complexity}.

\subsection{Memory allocation}
Apart from the numerical computation time, we are also interested in independently observing the memory allocation for both algorithms during the reconstruction process. For this reason, we record the memory usage every $0.2$ seconds. The evaluation for memory allocation is also performed independently of the investigation of numerical computation time in the previous section and separately for each algorithm. Similar to the numerical computation time, we present the evaluation for both increasing dimensions of scanning points and detector.
\begin{figure*}[!htb]
\centering
     \scalebox{0.42}{\input{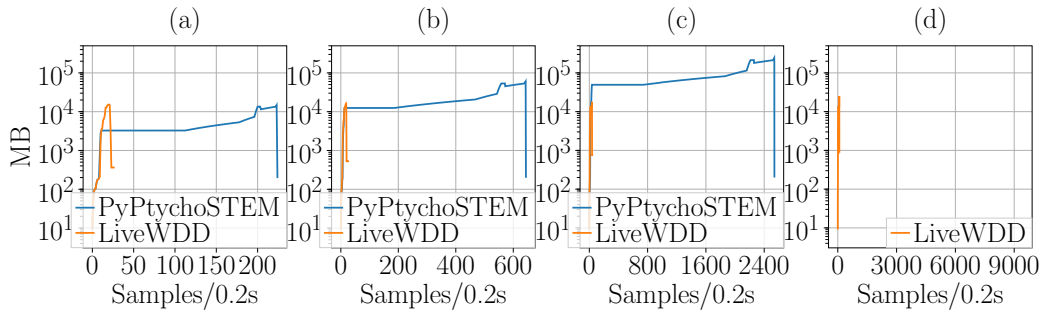}}  

\caption{Memory allocation for conventional \ac{WDD} and Live \ac{WDD} for different detector sizes: (a). (128,128,128,128), (b). (128,128,256,256), (c). (128,128,512,512), (d). (128,128,1024,1024). }
\label{fig:memory_increasing_detector} 
\end{figure*}

Figure \ref{fig:memory_increasing_detector} shows that for detector dimension $(1024,\\1024)$ PyPtychoSTEM requires more memory than available to complete the reconstruction. Live \ac{WDD} only requires a constant amount of memory around $10^4$ MB independent of detector size.
\begin{table}[htpb]
\centering
\caption{Maximum memory allocation in MB for PyPtychoSTEM and Live \ac{WDD} for fixed scanning points dimension $S_y = S_x = 128$ and increasing detector size $N_y = N_x$.}
\label{tab:memory_inc_det}
\begin{tabular*}{0.45\textwidth}{Y @{\extracolsep{\fill}} YYY}
\hline
    \multicolumn{1}{c}{Dimension} & \multicolumn{1}{c }{PyPtychoSTEM} & \multicolumn{1}{c}{Live WDD} \\\midrule
     128 &   14725 &  15343  \\
     256 &   60505 &  16136 \\
     512 &   244557 & 17889  \\
    1024 &   -      & 24887  \\
    \bottomrule
\end{tabular*}
\end{table}
The maximum memory allocation to complete each algorithm for increasing detector dimension is presented in Table \ref{tab:memory_inc_det}.
\begin{figure*}[!htb]
\centering
     \scalebox{0.42}{\input{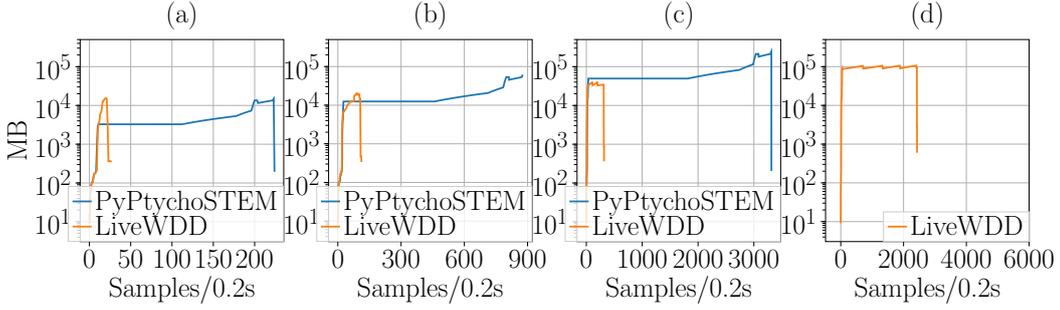}}  

\caption{Memory allocation for conventional \ac{WDD} and Live \ac{WDD} for different scanning point dimensions: (a).(128,128,128,128), (b).(256,256,128,128), (c).(512,512,128,128), (d).(1024,1024,128,128)}
\label{fig:memory_increasing_scan} 
\end{figure*}
In addition, we also investigate the effect of increasing scanning points dimension or the dimension of the field of view, as presented in Figure \ref{fig:memory_increasing_scan}. PyPtychoSTEM requires more memory than available to complete the reconstruction for scanning points $(1024,1024)$. Also in this scenario, Live \ac{WDD} uses memory more efficiently than PyPtychoSTEM. The maximum memory allocation for increasing scanning points is given in Table \ref{tab:memory_inc_scan}. 
\vspace*{-0.2cm}
\begin{table}[htpb]
\centering
\caption{Maximum memory allocation in MB for PyPtychoSTEM and Live \ac{WDD} for fixed dimension detector $N_y = N_x = 128$ and increasing scanning points $S_y = S_x$.}
\label{tab:memory_inc_scan}
\begin{tabular*}{0.45\textwidth}{Y @{\extracolsep{\fill}} YYY}
\hline
    \multicolumn{1}{c}{Dimension} & \multicolumn{1}{c }{PyPtychoSTEM} & \multicolumn{1}{c}{Live WDD} \\\midrule
     128 &   14725  &  15343  \\
     256 &   61521  &  19952 \\
     512 &   244121  & 39577  \\
    1024 &   -       & 106585  \\
    \bottomrule
\end{tabular*}
\end{table}
\subsection{Live processing evaluation}
In this section, we demonstrate that the performance of Live \ac{WDD} is sufficient for live acquisition and reconstruction with real-world 4D STEM detectors. In 4D \ac{STEM}, illustrated in Figure \ref{Fig:Ptycho}, the acquisition time per scanning point is usually limited by the detector frame rate. The specification for different detectors, namely Merlin MedipixEM \footnote{\url{https://quantumdetectors.com/wp-content/uploads/2022/01/MerlinEM-app-notes.pdf}}, Dectris Quadro\footnote{\url{https://www.dectris.com/detectors/electron-detectors/for-materials-science/quadro/}}, and Dectris Arina\footnote{\url{https://www.dectris.com/detectors/electron-detectors/for-materials-science/arina/}}, are given in Table \ref{tab:detector}. The maximum frame rate may depend on the chosen bit depth and readout area for a given detector.
\begin{table}[!htb]
  \caption{Specification of different detectors that support experimental acquisition}
\label{tab:detector}
 \centering
  \begin{tabular}{lll}
    \toprule
     \bfseries Detectors & \bfseries Frame rate (kHz) \\
     \bottomrule
      MedipixEM & 18.8 (1-bit), 3.2 (6-bit), or 1.6 (12-bit)\\
      
      Dectris Quadro &  	
2.25 (16-bit), 4.5 (8-bit),  ROI 9 (16-bit), \\ & ROI 18 (8-bit)
 \\
      Dectris Arina & 120 \\
    \bottomrule
  \end{tabular}
\end{table}
To accomplish a continuous reconstruction, data processing time per detector frame needs to be faster than the \ac{STEM} dwell time. Following the considerations on computational complexity, this highly depends on the number of scanning points, the detector size, and the number of non-zero entries in the Wiener filter. Since the processing is parallelized using a \ac{UDF} and LiberTEM-live, the number and speed of CPU cores is a major factor as well.

\begin{figure}[!htb]
\centering
     \scalebox{0.7}{
\pgfplotsset{every tick label/.append style={font=\LARGE}}
\begin{tikzpicture}

\definecolor{color0}{rgb}{0.12156862745098,0.466666666666667,0.705882352941177}

\begin{axis}[
width=5in,
height=2.5in,
tick align=outside,
tick pos=left,
x grid style={white!69.0196078431373!black},
xlabel={\LARGE Number of cores},
xmajorgrids,
xmin=-0.55, xmax=33.55,
xtick style={color=black},
y grid style={white!69.0196078431373!black},
ylabel={\LARGE Seconds},
ytick = {1,2,4,8,16, 32},
yticklabels = {1,2,4,8,16, 32},
ymajorgrids,
xmode = log,
ymode = log,
ymin=0, ymax=20,
ytick style={color=black},
xtick={1,2,4,8,16,32},
xticklabels={1,2,4,8, 16, 32},
]
\addplot [semithick, color0, mark=*, mark size=3, mark options={solid}]
table {%
1 18.3792677586898
2 9.33362670056522
4 4.92551750969142
8 2.69242353364825
16 1.66026053391397
32 1.23018423561007
};\addlegendentry{\LARGE Numerical Result}
\addplot [semithick, red, mark=*, mark size=3, mark options={solid}]
table {%
1 18.3792677586898
2 9.18963387934 
4 4.59481693967
8 2.29740846984
16 1.14870423492
32 0.57435211745
};
\addlegendentry{\LARGE Linear Reference}
\end{axis}

\end{tikzpicture}}  

\caption{Computation time depending on the number of cores for a dataset dimension $(128,128,128,128)$.
Despite using the CPU bandwidth as a shared resource the algorithm exhibits a $15\times$ speedup on 32 cores.}
\label{fig:no_cores}
\end{figure}
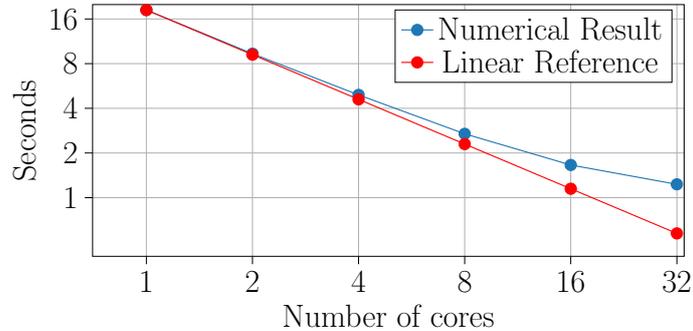

To illustrate the scalability of Live \ac{WDD}, we show the scaling behavior of the computation time as a function of number of CPU cores for dimension $\left(128,128,128,128\right)$, as presented in Figure \ref{fig:no_cores}. The scaling is nearly linear for up to eight cores and tapers off after that.
\begin{table}[htpb]
\centering
\caption{Processing speed in frames per second for Live \ac{WDD} for fixed dimension detector $N_y = N_x = 128$ and increasing scanning points $S_y = S_x$.}
\label{tab:fps_inc_scan}
\begin{tabular*}{0.45\textwidth}{Y @{\extracolsep{\fill}} YYY}
\hline
    \multicolumn{1}{c}{Dimension} & \multicolumn{1}{c }{Average (sec)} & \multicolumn{1}{c}{ Frame/sec} \\\midrule
     128 &   1.32  & 12412   \\
     256 &   14.82  &  4422 \\
     512 &   228.65  &  1146 \\
    1024 &   3048.66  & 343  \\
    \bottomrule
\end{tabular*}
\end{table} 
 
Based on the performance data from our $32$ core CPU in Tables \ref{tab:numeric_inc_det} and \ref{tab:numeric_inc_scan}, we can therefore support live reconstruction up to the frames per second (fps) as presented in Table \ref{tab:fps_inc_scan}. Therefore, a reconstruction using Live \ac{WDD} can keep up with  Merlin Medipix and Dectris Quadro without ROI up to a dimension of $\left(256,256,128,128\right)$ when used with the given setup and settings.

\begin{figure*}[!htb]
\centering
     \scalebox{1.}{
\begin{tikzpicture}

\begin{groupplot}[group style={group size=5 by 2,horizontal sep=0.5em, vertical sep=4.5em},width = 0.25*\textwidth, height=(0.25)*\textwidth]
\nextgroupplot[
colorbar horizontal,
colormap/blackwhite,
ticks = none,
title = \Large (a),
ylabel = \Large Live \ac{WDD},
point meta max=0.251606047153473,
point meta min=-0.261222332715988,
tick align=outside,
tick pos=left,
x grid style={white!69.0196078431373!black},
xmin=-0.5, xmax=127.5,
xtick style={color=black},
y dir=reverse,
y grid style={white!69.0196078431373!black},
ymin=-0.5, ymax=127.5,
ytick style={color=black}
]
\addplot graphics [includegraphics cmd=\pgfimage,xmin=-0.5, xmax=127.5, ymin=127.5, ymax=-0.5] {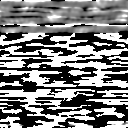};

\nextgroupplot[
colorbar horizontal,
colormap/blackwhite,
hide y axis,
ticks = none,
point meta max=0.333269238471985,
point meta min=-0.273211002349854,
title = \Large (b),
tick align=outside,
tick pos=left,
x grid style={white!69.0196078431373!black},
xmin=-0.5, xmax=127.5,
xtick style={color=black},
y dir=reverse,
y grid style={white!69.0196078431373!black},
ymin=-0.5, ymax=127.5,
ytick style={color=black}
]
\addplot graphics [includegraphics cmd=\pgfimage,xmin=-0.5, xmax=127.5, ymin=127.5, ymax=-0.5] {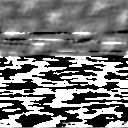};

\nextgroupplot[
colorbar horizontal,
colormap/blackwhite,
hide y axis,
ticks = none,
title = \Large (c),
point meta max=0.331693083047867,
point meta min=-0.285140573978424,
tick align=outside,
tick pos=left,
x grid style={white!69.0196078431373!black},
xmin=-0.5, xmax=127.5,
xtick style={color=black},
y dir=reverse,
y grid style={white!69.0196078431373!black},
ymin=-0.5, ymax=127.5,
ytick style={color=black}
]
\addplot graphics [includegraphics cmd=\pgfimage,xmin=-0.5, xmax=127.5, ymin=127.5, ymax=-0.5] {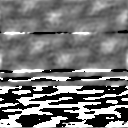};

\nextgroupplot[
colorbar horizontal,
colormap/blackwhite,
hide y axis,
title = \Large (d),
ticks = none,
point meta max=0.261677473783493,
point meta min=-0.184297353029251,
tick align=outside,
tick pos=left,
x grid style={white!69.0196078431373!black},
xmin=-0.5, xmax=127.5,
xtick style={color=black},
y dir=reverse,
y grid style={white!69.0196078431373!black},
ymin=-0.5, ymax=127.5,
ytick style={color=black}
]
\addplot graphics [includegraphics cmd=\pgfimage,xmin=-0.5, xmax=127.5, ymin=127.5, ymax=-0.5] {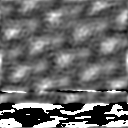};

\nextgroupplot[
colorbar horizontal,
colormap/blackwhite,
ticks = none,
ylabel = \Large Scan y,
title = \Large (e),
point meta max=0.263222128152847,
point meta min=-0.165853321552277,
tick align=outside,
tick pos=left,
x grid style={white!69.0196078431373!black},
xmin=-0.5, xmax=127.5,
xtick style={color=black},
y dir=reverse,
yticklabel pos=right,
y grid style={white!69.0196078431373!black},
ymin=-0.5, ymax=127.5,
ytick style={color=black}
]
\addplot graphics [includegraphics cmd=\pgfimage,xmin=-0.5, xmax=127.5, ymin=127.5, ymax=-0.5] {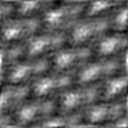};

\nextgroupplot[
colorbar horizontal,
colormap/blackwhite,
ticks = none,
ylabel = \Large Live \ac{SSB},
xlabel = \Large Scan x,
point meta max=0.810894383465788,
point meta min=-0.798732729192162,
tick align=outside,
tick pos=left,
x grid style={white!69.0196078431373!black},
xmin=-0.5, xmax=127.5,
xtick style={color=black},
y dir=reverse,
y grid style={white!69.0196078431373!black},
ymin=-0.5, ymax=127.5,
ytick style={color=black}
]
\addplot graphics [includegraphics cmd=\pgfimage,xmin=-0.5, xmax=127.5, ymin=127.5, ymax=-0.5] {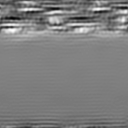};

\nextgroupplot[
colorbar horizontal,
colormap/blackwhite,
hide y axis,
ticks = none,
xlabel = \Large Scan x,
point meta max=0.521422318986998,
point meta min=-0.319312126020433,
tick align=outside,
tick pos=left,
x grid style={white!69.0196078431373!black},
xmin=-0.5, xmax=127.5,
xtick style={color=black},
y dir=reverse,
y grid style={white!69.0196078431373!black},
ymin=-0.5, ymax=127.5,
ytick style={color=black}
]
\addplot graphics [includegraphics cmd=\pgfimage,xmin=-0.5, xmax=127.5, ymin=127.5, ymax=-0.5] {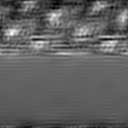};

\nextgroupplot[
colorbar horizontal,
colormap/blackwhite,
hide y axis,
ticks = none,
xlabel = \Large Scan x,
point meta max=0.360626965473703,
point meta min=-0.211943005959824,
tick align=outside,
tick pos=left,
x grid style={white!69.0196078431373!black},
xmin=-0.5, xmax=127.5,
xtick style={color=black},
y dir=reverse,
y grid style={white!69.0196078431373!black},
ymin=-0.5, ymax=127.5,
ytick style={color=black}
]
\addplot graphics [includegraphics cmd=\pgfimage,xmin=-0.5, xmax=127.5, ymin=127.5, ymax=-0.5] {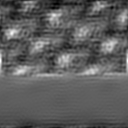};

\nextgroupplot[
colorbar horizontal,
colormap/blackwhite,
hide y axis,
ticks = none,
xlabel = \Large Scan x,
point meta max=0.262921147093797,
point meta min=-0.158436343361057,
tick align=outside,
tick pos=left,
x grid style={white!69.0196078431373!black},
xmin=-0.5, xmax=127.5,
xtick style={color=black},
y dir=reverse,
y grid style={white!69.0196078431373!black},
ymin=-0.5, ymax=127.5,
ytick style={color=black}
]
\addplot graphics [includegraphics cmd=\pgfimage,xmin=-0.5, xmax=127.5, ymin=127.5, ymax=-0.5] {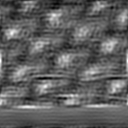};

\nextgroupplot[
colorbar horizontal,
colormap/blackwhite,
ticks = none,
xlabel = \Large Scan x,
ylabel = \Large Scan y,
yticklabel pos=right,
point meta max=0.208665738235203,
point meta min=-0.106166752151338,
tick align=outside,
tick pos=left,
x grid style={white!69.0196078431373!black},
xmin=-0.5, xmax=127.5,
xtick style={color=black},
y dir=reverse,
y grid style={white!69.0196078431373!black},
ymin=-0.5, ymax=127.5,
ytick style={color=black}
]
\addplot graphics [includegraphics cmd=\pgfimage,xmin=-0.5, xmax=127.5, ymin=127.5, ymax=-0.5] {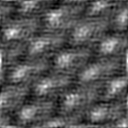};

\end{groupplot}
\end{tikzpicture}}  

\caption{Snapshots of simulated live ptychography (phase) using Live \ac{WDD} (Top) and Live \ac{SSB} (Bottom) method at different stages of the scan for \ac{SrTiO$_3$} datasets for different buffer processes: (a) 10, (b) 26, (c) 41, (d) 57, (e) 72. Note how the results don't have to be merged in any particular order.}
\label{fig:live_update} 
\end{figure*}

Figure \ref{fig:live_update} shows simulated live reconstruction for different stages of scanning progress for Live \ac{WDD} and Live \ac{SSB} from \citep{Strauch2021}, where the update is added gradually until completing all scanning points. Here the dimension of the four-dimensional \ac{STEM} data is\\ $\left(128,128,256,256\right)$, as described in Table \ref{tab:params_exp}.
 
 \section{Discussion}
Since we only perform the object reconstruction using a synthetic probe initialization, a potential next step for Live WDD could be to factor in the microscope alignment and update the probe. After estimating the object, we can swap the deconvolution process to reconstruct the probe. This process is straightforward, but the additional computation and update for the Wiener filter would affect the performance. Therefore, an efficient and on-fly computation should be implemented to overcome this issue.

The time complexity of Live \ac{WDD} for a large field of view is scaled quadratically with total scanning points and reduced dimension, respectively. Although the Live \ac{WDD} implementation can complete the reconstruction, our numerical observation shows that it takes approximately one hour with dimension $(1024,1024,128,128)$, which is inefficient. This could be overcome by subdividing the field of view into smaller patches that are reconstructed independently and subsequently merged. In this case, the time complexity can be reduced to a linear scale of the number of subsets.

Another strategy to optimize the performance of Live \ac{WDD} could be to choose an optimal scan step based on the intersection of the probe's auto correlation in reciprocal space, as presented in \eqref{eq:low_correlation}. Making sure that the scan grid is not unnecessarily fine can reducing the number of scanning points to process.

It is also thinkable to adapt \ac{WDD} and Live \ac{WDD} for scan patterns that are not on an equispaced grid. In that case a matrix for a non-uniform discrete Fourier transform should be used to match the scanning points. We will defer such possible improvements to future works.

\section{Summary}

\textcolor{black}{As an evolution of the classical \ac{WDD} algorithm, we demonstrated Live \ac{WDD} that can reconstruct in a streaming fashion while acquiring diffraction patterns to support real-time reconstruction. Our investigation shows that Live \ac{WDD} produces object reconstructions that approximate the conventional result. The algorithm uses less memory and runs faster than the classical \ac{WDD} for typical parameters.} 
\textcolor{black}{As a side effect of dimensionality reduction, we also observe that it acts as a filter for Poissonian noise to attain a more robust reconstruction from low-dose diffraction patterns. We compare the numerical computation time of the proposed algorithm with the dwell time of Merlin Medipix and Dectris Quadro detectors, where we can perform live continuous reconstruction with a field of view up to $(256, 256)$ on a system with 32 CPU cores.}
 
\section*{Acknowledgements}
Arya Bangun, Alexander Clausen, Dieter Weber, Rafal E. Dunin-Borkowski acknowledge support from Helmholtz Association under contract No.~ZT-I-0025 (Ptychography 4.0) and  JL-MDMC (Joint Lab on Model and Data-Driven Material Characterization). Paul  F. Baumeister acknowledges the support from SiVeGCS (Sicherstellung der weiteren Verfügbarkeit der Supercomputing-Ressourcen des GCS).

\bibliographystyle{MandM}
\bibliography{references}

\end{document}